\documentclass[10pt,final]{IEEEtran}
\IEEEoverridecommandlockouts

\usepackage{amssymb}
\usepackage{amsmath}
\usepackage[dvips]{graphicx}
\usepackage{color}
\usepackage{amsthm}
\usepackage{hyperref}
\usepackage{url}
\usepackage{cite}
\usepackage[nolist,printonlyused]{acronym}
\usepackage{capt-of}
\usepackage[labelformat=simple]{subcaption}

\usepackage{acronym}
\def\appearancecount{3}
\let\oldacronymused\acronymused
\AtBeginDocument{\let\acronymused\oldacronymused}
\makeatletter
\renewcommand*{\acronymused}[1]{%
    \expandafter\ifx\csname acused@#1\endcsname\AC@used
        \expandafter\xdef\csname
acusedcount@#1\endcsname{\the\numexpr1+\csname
acusedcount@#1\endcsname}
    \else
        \global\expandafter\let\csname acused@#1\endcsname\AC@used
        \global\let\AC@populated\AC@used
        \expandafter\gdef\csname acusedcount@#1\endcsname{1}
    \fi
}

\def\AC@iffewuses#1{%
    \expandafter\ifx\csname acusedcount@#1\endcsname\relax
        \expandafter\@firstoftwo
    \else
        \ifnum\csname acusedcount@#1\endcsname<\appearancecount\relax
            \expandafter\@firstoftwo\romannumeral-`0%
        \else
            \expandafter\@secondoftwo\romannumeral-`0%
        \fi
    \fi
}

\let\@@ac\@ac
\def\@ac#1{\AC@iffewuses{#1}{\ifAC@starred\acl*{#1}\else\acl{#1}\fi}{\@@ac{#1}}}

\let\@@acp\@acp
\def\@acp#1{\AC@iffewuses{#1}{\ifAC@starred\aclp*{#1}\else\aclp{#1}\fi}{\@@acp{#1}}}

\usepackage{pgf}
\usepackage{pgfplots}
\usepackage{tikz}
\usepackage{grffile}
\pgfplotsset{compat=newest}
\usetikzlibrary{plotmarks}
\usepgfplotslibrary{patchplots}
\usetikzlibrary{quotes,angles,automata,arrows,positioning,calc,matrix,decorations.markings,decorations.pathreplacing,patterns,arrows.meta}
 \pgfdeclarepatternformonly{south west lines}{\pgfqpoint{-0pt}{-0pt}}{\pgfqpoint{3pt}{3pt}}{\pgfqpoint{3pt}{3pt}}{
        \pgfsetlinewidth{0.4pt}
        \pgfpathmoveto{\pgfqpoint{0pt}{0pt}}
        \pgfpathlineto{\pgfqpoint{3pt}{3pt}}
        \pgfpathmoveto{\pgfqpoint{2.8pt}{-.2pt}}
        \pgfpathlineto{\pgfqpoint{3.2pt}{.2pt}}
        \pgfpathmoveto{\pgfqpoint{-.2pt}{2.8pt}}
        \pgfpathlineto{\pgfqpoint{.2pt}{3.2pt}}
        \pgfusepath{stroke}}
\makeatletter
\newcommand{\vast}{\bBigg@{3}}
\newcommand{\Vast}{\bBigg@{4}}
\makeatother

\usepackage{multirow}
\usepackage{makecell}
\newif\ifthickvlines
\newcolumntype{|}{!{%
  \ifthickvlines
    \vrule width2\arrayrulewidth
  \else
    \vline
  \fi}}

\newcommand{\etal}[1]{#1~{\textit{et al.}}}

\begin{acronym}[LTE-Advanced]
  \acro{2G}{Second Generation}
  \acro{3-DAP}{3-Dimensional Assignment Problem}
  \acro{3G}{3$^\text{rd}$~Generation}
  \acro{3GPP}{3$^\text{rd}$~Generation Partnership Project}
  \acro{4G}{4$^\text{th}$~Generation}
  \acro{5G}{5$^\text{th}$~Generation}
  \acro{AA}{Antenna Array}
  \acro{AC}{Admission Control}
  \acro{AD}{Attack-Decay}
  \acro{ADSL}{Asymmetric Digital Subscriber Line}
  \acro{AHW}{Alternate Hop-and-Wait}
  \acro{AMC}{Adaptive Modulation and Coding}
  \acro{AoI}{Age of Information}
  \acro{AP}{access point}
  \acro{APA}{Adaptive Power Allocation}
  \acro{AR}{Augmented reality}
  \acro{ARMA}{Autoregressive Moving Average}
  \acro{ATES}{Adaptive Throughput-based Efficiency-Satisfaction Trade-Off}
  \acro{AWGN}{Additive White Gaussian Noise}
  \acro{BB}{Branch and Bound}
  \acro{BBU}{Baseband Unit}
  \acro{BD}{Block Diagonalization}
  \acro{BER}{Bit Error Rate}
  \acro{BF}{Best Fit}
  \acro{BFD}{Bidirectional Full Duplex}
  \acro{BGP}{Border Gateway Protocol}
  \acro{BLER}{BLock Error Rate}
  \acro{BP}{Back-pressure}
  \acro{BPC}{Binary Power Control}
  \acro{BPSK}{Binary Phase-Shift Keying}
  \acro{BRA}{Balanced Random Allocation}
  \acro{BS}{base station}
  \acro{CAP}{Combinatorial Allocation Problem}
  \acro{CAPEX}{Capital Expenditure}
  \acro{CBF}{Coordinated Beamforming}
  \acro{CBR}{Constant Bit Rate}
  \acro{CBS}{Class Based Scheduling}
  \acro{CC}{Congestion Control}
  \acro{CCN}{Content-Centric Networking}
  \acro{CDF}{Cumulative Distribution Function}
  \acro{CDMA}{Code-Division Multiple Access}
  \acro{CL}{Closed Loop}
  \acro{CLPC}{Closed Loop Power Control}
  \acro{CNR}{Channel-to-Noise Ratio}
  \acro{CoDel}{Controlled Delay}
  \acro{CPA}{Cellular Protection Algorithm}
  \acro{CPICH}{Common Pilot Channel}
  \acro{CoMP}{Coordinated Multi-Point}
  \acro{CP}{Cyclic Prefix}
  \acro{CQI}{Channel Quality Indicator}
  \acro{CRM}{Constrained Rate Maximization}
	\acro{CRN}{Cognitive Radio Network}
  \acro{CS}{Coordinated Scheduling}
  \acro{CSI}{channel state information}
  \acro{CSMA}{Carrier-Sense Multiple Access}
  \acro{CSMA/CA}{Carrier-Sense Multiple Access with Collision Avoidance}
  \acro{CTS}{Clear to Send}
  \acro{CUE}{Cellular User Equipment}
  \acro{D2D}{Device-to-Device}
  \acro{DCA}{Dynamic Channel Allocation}
  \acro{DE}{Differential Evolution}
  \acro{DFT}{Discrete Fourier Transform}
  \acro{DIST}{Distance}
  \acro{DL}{Downlink}
  \acro{DMA}{Double Moving Average}
  \acro{DMRS}{Demodulation Reference Signal}
  \acro{D2DM}{D2D Mode}
  \acro{DMS}{D2D Mode Selection}
  \acro{DPC}{Dirty Paper Coding}
  \acro{DRA}{Dynamic Resource Assignment}
  \acro{DSA}{Dynamic Spectrum Access}
  \acro{DSM}{Delay-based Satisfaction Maximization}
  \acro{ECC}{Electronic Communications Committee}
  \acro{EFLC}{Error Feedback Based Load Control}
  \acro{EI}{Efficiency Indicator}
  \acro{eNB}{Evolved Node B}
  \acro{EPA}{Equal Power Allocation}
  \acro{EPC}{Evolved Packet Core}
  \acro{EPS}{Evolved Packet System}
  \acro{E-UTRAN}{Evolved Universal Terrestrial Radio Access Network}
  \acro{ES}{Exhaustive Search}
  \acro{FBMC}{Filter Bank Multi-Carrier}
  \acro{FD}{Full Duplex}
  \acro{FDD}{Frequency Division Duplex}
  \acro{FDM}{Frequency Division Multiplexing}
  \acro{FDMA}{Frequency Division Multiple Access}
  \acro{FER}{Frame Erasure Rate}
  \acro{FF}{Fast Fading}
  \acro{FIB}{Forwarding Information Base}
  \acro{FIFO}{First In First Out}
  \acro{FL}{Fast-Lipschitz}
  \acro{F-OFDM}{Filter-OFDM}
  \acro{FSB}{Fixed Switched Beamforming}
  \acro{FST}{Fixed SNR Target}
  \acro{FTP}{File Transfer Protocol}
  \acro{GA}{Genetic Algorithm}
  \acro{GBR}{Guaranteed Bit Rate}
  \acro{GFDM}{Generalized Frequency Division Multiplexing}
  \acro{GLR}{Gain to Leakage Ratio}
  \acro{GOS}{Generated Orthogonal Sequence}
  \acro{GPL}{GNU General Public License}
  \acro{GQC}{General Qualifying Conditions}
  \acro{GRP}{Grouping}
  \acro{GSM}{Global System for Mobile}
  \acro{ARQ}{Automatic Repeat Request}
  \acro{HARQ}{Hybrid Automatic Repeat Request}
  \acro{HD}{Half-Duplex}
  \acro{HMS}{Harmonic Mode Selection}
  \acro{HOL}{Head Of Line}
  \acro{HSDPA}{High-Speed Downlink Packet Access}
  \acro{HSPA}{High Speed Packet Access}
  \acro{HTTP}{HyperText Transfer Protocol}
  \acro{HULL}{High-bandwidth Ultra-low Latency}
  \acro{ICMP}{Internet Control Message Protocol}
  \acro{ICI}{Intercell Interference}
  \acro{ID}{Identification}
  \acro{IETF}{Internet Engineering Task Force}
  \acro{ILP}{Integer Linear Program}
  \acro{JRAPAP}{Joint RB Assignment and Power Allocation Problem}
  \acro{UID}{Unique Identification}
  \acro{ICT}{information and communication technology}
  \acro{IID}{Independent and Identically Distributed}
  \acro{IIR}{Infinite Impulse Response}
  \acro{ILP}{Integer Linear Problem}
  \acro{IMT}{International Mobile Telecommunications}
  \acro{INV}{Inverted Norm-based Grouping}
  \acro{IoT}{Internet of Things}
  \acro{IP}{Integer Programming}
  \acro{IPv6}{Internet Protocol Version 6}
  \acro{ISD}{Inter-Site Distance}
  \acro{ISI}{Inter Symbol Interference}
  \acro{ITU}{International Telecommunication Union}
  \acro{JAFM}{Joint Assignment and Fairness Maximization}
  \acro{JASEM}{Joint Assignment and Spectral Efficiency Maximization}
  \acro{JOAS}{Joint Opportunistic Assignment and Scheduling}
  \acro{JOS}{Joint Opportunistic Scheduling}
  \acro{JP}{Joint Processing}
	\acro{JS}{Jump-Stay}
  \acro{KKT}{Karush-Kuhn-Tucker}
  \acro{L3}{Layer-3}
  \acro{LAC}{Link Admission Control}
  \acro{LA}{Link Adaptation}
  \acro{LAA}{License Assisted Access}
  \acro{LED}{Light Emitting Diode}
  \acro{LC}{Load Control}
  \acro{LOS}{Line of Sight}
  \acro{LP}{Linear Programming}
  \acro{LTE}{Long Term Evolution}
	\acro{LTE-A}{\ac{LTE}-Advanced}
  \acro{LTE-Advanced}{Long Term Evolution Advanced}
  \acro{LTE-U}{LTE in Unlicensed Spectrum}
  \acro{M2M}{Machine-to-Machine}
  \acro{MAC}{Medium Access Control}
  \acro{MANET}{Mobile Ad hoc Network}
  \acro{MC}{multi-carrier}
  \acro{MC-NOMA}{Multi-Carrier - Non-Orthogonal Multiple Access}
  \acro{MCS}{Modulation and Coding Scheme}
  \acro{MDB}{Measured Delay Based}
  \acro{MDI}{Minimum D2D Interference}
  \acro{MF}{Matched Filter}
  \acro{MFH}{Mobile Fronthaul}
  \acro{MG}{Maximum Gain}
  \acro{MH}{Multi-Hop}
  \acro{MIMO}{Multiple Input Multiple Output}
  \acro{MINLP}{Mixed Integer Nonlinear Programming}
  \acro{MIP}{Mixed Integer Programming}
  \acro{MISO}{Multiple Input Single Output}
  \acro{MLWDF}{Modified Largest Weighted Delay First}
  \acro{MME}{Mobility Management Entity}
  \acro{MMSE}{Minimum Mean Square Error}
  \acro{mmWave}{Millimeter-Wave}
  \acro{MOS}{Mean Opinion Score}
  \acro{MPF}{Multicarrier Proportional Fair}
  \acro{MRA}{Maximum Rate Allocation}
  \acro{MR}{Maximum Rate}
  \acro{MRC}{Maximum Ratio Combining}
  \acro{MRT}{Maximum Ratio Transmission}
  \acro{MRUS}{Maximum Rate with User Satisfaction}
  \acro{MS}{Mode Selection}
  \acro{MSE}{Mean Squared Error}
  \acro{MSI}{Multi-Stream Interference}
  \acro{MTC}{Machine-Type Communication}
  \acro{MTSI}{Multimedia Telephony Services over IMS}
  \acro{MTSM}{Modified Throughput-based Satisfaction Maximization}
  \acro{MU-MIMO}{Multi-User Multiple Input Multiple Output}
  \acro{MU}{Multi-User}
  \acro{NAS}{Non-Access Stratum}
  \acro{NB}{Node B}
	\acro{NCL}{Neighbor Cell List}
  \acro{NFV}{Network Function Virtualization}
  \acro{NLP}{Nonlinear Programming}
  \acro{NLOS}{Non-Line of Sight}
  \acro{NMSE}{Normalized Mean Square Error}
  \acro{NOMA}{Non-Orthogonal Multiple Access}
  \acro{NORM}{Normalized Projection-based Grouping}
  \acro{NP}{Non-deterministic Polynomial-time}
  \acro{NRT}{Non-Real Time}
  \acro{NSPS}{National Security and Public Safety Services}
  \acro{O2I}{Outdoor to Indoor}
  \acro{OFDMA}{Orthogonal Frequency Division Multiple Access}
  \acro{OFDM}{Orthogonal Frequency Division Multiplexing}
  \acro{OFPC}{Open Loop with Fractional Path Loss Compensation}
	\acro{O2I}{Outdoor-to-Indoor}
  \acro{OL}{Open Loop}
  \acro{OLPC}{Open-Loop Power Control}
  \acro{OL-PC}{Open-Loop Power Control}
  \acro{OMA}{Orthogonal Multiple Access}
  \acro{OPEX}{Operational Expenditure}
  \acro{OSPF}{Open Shortest Path}
  \acro{ORB}{Orthogonal Random Beamforming}
  \acro{JO-PF}{Joint Opportunistic Proportional Fair}
  \acro{OSI}{Open Systems Interconnection}
  \acro{PAIR}{D2D Pair Gain-based Grouping}
  \acro{PAPR}{Peak-to-Average Power Ratio}
  \acro{P2P}{Peer-to-Peer}
  \acro{PC}{Power Control}
  \acro{PCI}{Physical Cell ID}
  \acro{PHY}{Physical Layer}
  \acro{PDCCH}{Physical Downlink Control Channel}
  \acro{PDF}{Probability Density Function}
  \acro{PER}{Packet Error Rate}
  \acro{PF}{Proportional Fair}
  \acro{P-GW}{Packet Data Network Gateway}
  \acro{PIB}{Pending Interest Table}
  \acro{PL}{Pathloss}
  \acro{PRB}{Physical Resource Block}
  \acro{PROJ}{Projection-based Grouping}
  \acro{ProSe}{Proximity Services}
  \acro{PS}{Packet Scheduling}
  \acro{PSO}{Particle Swarm Optimization}
  \acro{PSTN}{Public Switched Telephone Network}
  \acro{PUCCH}{Physical Uplink Control Channel}
  \acro{PZF}{Projected Zero-Forcing}
  \acro{QAM}{Quadrature Amplitude Modulation}
  \acro{QoS}{Quality of Service}
  \acro{QPSK}{Quadri-Phase Shift Keying}
  \acro{RAISES}{Reallocation-based Assignment for Improved Spectral Efficiency and Satisfaction}
  \acro{RAN}{Radio Access Network}
  \acro{RA}{Resource Allocation}
  \acro{RAT}{Radio Access Technology}
  \acro{RATE}{Rate-based}
  \acro{RB}{Resource Block}
  \acro{RBG}{Resource Block Group}
  \acro{REF}{Reference Grouping}
  \acro{RF}{Radio-Frequency}
  \acro{RFID}{radio-frequency identification}
  \acro{RLC}{Radio Link Control}
  \acro{RM}{Rate Maximization}
  \acro{RNC}{Radio Network Controller}
  \acro{RND}{Random Grouping}
  \acro{RRA}{Radio Resource Allocation}
  \acro{RRH}{Remote Radio Heads}
  \acro{RRM}{Radio Resource Management}
  \acro{RSCP}{Received Signal Code Power}
  \acro{RSRP}{Reference Signal Receive Power}
  \acro{RSRQ}{Reference Signal Receive Quality}
  \acro{RR}{Round Robin}
  \acro{RRC}{Radio Resource Control}
  \acro{RSSI}{Received Signal Strength Indicator}
  \acro{RT}{Real Time}
  \acro{RTS}{Request to Send}
  \acro{RU}{Resource Unit}
  \acro{RUNE}{RUdimentary Network Emulator}
  \acro{RV}{Random Variable}
  \acro{SAC}{Session Admission Control}
  \acro{SCM}{Spatial Channel Model}
  \acro{SCMA}{Sparse Code Multiple Access}
  \acro{SC}{Single-carrier}
  \acro{SC-FDMA}{Single Carrier - Frequency Division Multiple Access}
  \acro{SC-NOMA}{Single Carrier - Non-Orthogonal Multiple Access}
  \acro{SD}{Soft Dropping}
  \acro{S-D}{Source-Destination}
  \acro{SDPC}{Soft Dropping Power Control}
  \acro{SDMA}{Space-Division Multiple Access}
  \acro{SDN}{Software-Defined Networking}
  \acro{SER}{Symbol Error Rate}
  \acro{SES}{Simple Exponential Smoothing}
  \acro{S-GW}{Serving Gateway}
  \acro{SINR}{Signal-to-Interference-plus-Noise Ratio}
  \acro{SI}{Self-Interference}
  \acro{SIP}{Session Initiation Protocol}
  \acro{SISO}{Single Input Single Output}
  \acro{SIMO}{Single Input Multiple Output}
  \acro{SIR}{Signal to Interference Ratio}
  \acro{SLNR}{Signal-to-Leakage-plus-Noise Ratio}
  \acro{SMA}{Simple Moving Average}
  \acro{SNR}{signal to noise ratio}
  \acro{SORA}{Satisfaction Oriented Resource Allocation}
  \acro{SORA-NRT}{Satisfaction-Oriented Resource Allocation for Non-Real Time Services}
  \acro{SORA-RT}{Satisfaction-Oriented Resource Allocation for Real Time Services}
  \acro{SPF}{Single-Carrier Proportional Fair}
  \acro{SRA}{Sequential Removal Algorithm}
  \acro{SRS}{Sounding Reference Signal}
  \acro{SU-MIMO}{Single-User Multiple Input Multiple Output}
  \acro{SU}{Single-User}
  \acro{SVD}{Singular Value Decomposition}
  \acro{TCP}{Transmission Control Protocol}
  \acro{TDD}{Time Division Duplex}
  \acro{TDMA}{Time Division Multiple Access}
  \acro{TNFD}{Three Node Full-Duplex}
  \acro{TETRA}{Terrestrial Trunked Radio}
  \acro{TP}{Transmit Power}
  \acro{TPC}{Transmit Power Control}
  \acro{TSN}{Time-Sensitive Networking}
  \acro{TTI}{Transmission Time Interval}
  \acro{TTR}{Time-To-Rendezvous}
  \acro{TSM}{Throughput-based Satisfaction Maximization}
  \acro{TU}{Typical Urban}
  \acro{UDP}{User Datagram Protocol}
  \acro{UE}{User Equipment}
  \acro{UEPS}{Urgency and Efficiency-based Packet Scheduling}
  \acro{UFMC}{Universal Filtered Multi-Carrier}
  \acro{UL}{Uplink}
  \acro{UMTS}{Universal Mobile Telecommunications System}
  \acro{URI}{Uniform Resource Identifier}
  \acro{URLLC}{Ultra Reliable Low-Latency Communication}
  \acro{URM}{Unconstrained Rate Maximization}
  \acro{UMW}{Universal Max-Weight}
  \acro{VLC}{Visible Light Communication}
  \acro{VR}{Virtual reality}
  \acro{VoIP}{Voice over IP}
  \acro{WAN}{Wireless Access Network}
  \acro{WCDMA}{Wideband Code Division Multiple Access}
  \acro{WF}{Water-Filling}
  \acro{WiMAX}{Worldwide Interoperability for Microwave Access}
  \acro{WINNER}{Wireless World Initiative New Radio}
  \acro{WirelessHP}{Wireless Communication for High Performance}
  \acro{WLAN}{Wireless Local Area Network}
  \acro{WMPF}{Weighted Multicarrier Proportional Fair}
  \acro{WPF}{Weighted Proportional Fair}
  \acro{WSN}{Wireless Sensor Network}
  \acro{WWW}{World Wide Web}
  \acro{XIXO}{(Single or Multiple) Input (Single or Multiple) Output}
  \acro{ZF}{Zero-Forcing}
  \acro{ZMCSCG}{Zero Mean Circularly Symmetric Complex Gaussian}
\end{acronym}

\begin{document}

\title{Low-latency Networking: \\ Where  Latency Lurks and How to Tame It}
\author{Xiaolin Jiang, Hossein S. Ghadikolaei,~\IEEEmembership{Student Member,~IEEE,}
Gabor Fodor,~\IEEEmembership{Senior Member,~IEEE,} \\Eytan Modiano,~\IEEEmembership{Fellow,~IEEE,} Zhibo Pang,~\IEEEmembership{Senior Member,~IEEE,} Michele Zorzi,~\IEEEmembership{Fellow,~IEEE,} \\and Carlo Fischione~\IEEEmembership{Member,~IEEE}}

\maketitle

\begin{abstract}
While the current generation of mobile and fixed communication networks has been standardized for mobile broadband services, the next generation is driven by the vision of the Internet of Things and mission critical communication services requiring latency in the order of milliseconds or sub-milliseconds. However, these new stringent requirements have a large technical impact on the design of all layers of the communication protocol stack. The cross layer interactions are complex due to the multiple design principles and technologies that contribute to the layers' design and fundamental performance limitations. We will be able to develop low-latency networks only if we address the problem of these complex interactions from the new point of view of sub-milliseconds latency.
In this article, we propose a holistic analysis and classification of the main design principles and enabling technologies that will  make it possible to  deploy low-latency wireless communication networks.
We argue that these design principles and enabling technologies must be carefully orchestrated to meet the stringent requirements and to manage the inherent trade-offs between low latency and traditional performance metrics.
We also review currently ongoing standardization activities in prominent standards associations, and discuss open problems for future research.
\end{abstract}


\section{Introduction}
\label{Section: Introduction}

A series of major technological revolutions have pushed the development of communication networks to the current state-of-the-art that includes Internet, pervasive broadband wireless access and low data rate \ac{IoT}. One of the first crucial steps of this series of revolutions is the global \ac{PSTN}, the aggregate of the world's nationwide circuit switched telephone networks, which was designed to deliver arguably the most demanded and revenue generating services in the history of communication networks,
namely circuit switched voice.
As noted in~\cite{kelly1991network}, the \ac{PSTN} not only ensures the latency requirements imposed by voice communication services,
but also responds to randomly fluctuating demands and failures by rerouting traffic and reallocating communication resources.
Due to its ability to reliably meet human-centered latency requirements and deliver the popular voice service over very long distances,
even in the presence of fluctuating traffic demands and link failures, the \ac{PSTN} is a technology to which Mark Weiser's observation truly applies:
\begin{quote}
``The most profound technologies are those that disappear.
They weave themselves into the fabric of everyday life until they are indistinguishable from it."
~\cite{weiser1991computer,corson2010toward}
\end{quote}

The second step in the communication networks revolutions has made \ac{PSTN} indistinguishable from our everyday life. Such step is the \ac{GSM} communication standards suite. In the beginning of the 2000, \ac{GSM} has become the most widely spread mobile communications system, thanks to the support for users mobility, subscriber identity confidentiality, subscriber authentication as well as confidentiality of user traffic and signaling~\cite{katugampala2005real}.
The \ac{PSTN} and its extension via the \ac{GSM} wireless access networks have been a tremendous success in terms of Weiser's vision,
and also paved the way for new business models built around mobility,
high reliability, and latency as required from the perspective of voice services.

The success of \ac{PSTN} and \ac{GSM} was ultimately due to the development of packet switching and wireless access methods. The circuit switching technology dedicates communication resources along the end-to-end path of pairs of transmitters and receivers on a coarse time scale, which leads to poor resource utilization. In contrast, with packet switching, a communication network is designed to share a single link
between multiple pairs of transmitters and receivers. Although this idea boosts communication resource utilization, it may lead to packet congestion and consequent increase of latency. In fact, the Internet employs packet switching and \ac{VoIP} technologies, and the interval between the delivery of voice and data packets at the receiving end is not deterministic: packets can arrive out of order, have variable latencies (jitter) or even get lost. To devise mechanisms that help \ac{VoIP} systems to keep latency low and to compensate for jitter, \ac{VoIP} endpoints use buffers to delay incoming packets so as to create a steady stream. For voice services delivered by \ac{VoIP} technologies, the standard answer to the natural question ``how much latency is too much?" is dictated by the ability of humans to adapt the cadence of a conversation, and is typically quantified as around 200-250 ms.

The boundary of 200-250 ms has been further pushed by the third and fourth steps of the communication network revolutions. As more powerful wireless access generations succeeded \ac{GSM}, the 3rd Generation Partnership Project has designed the third and fourth generations of wireless cellular networks to meet 150 ms and subsequently an order of magnitude lower latency requirements in the wireless access part~\cite{chen2008voip}. However, these third and fourth generations of wireless cellular networks were mostly driven by the need of high data rates and coverage. The requirements of low-latency for mission critical applications, such as tele-surgery, virtual reality over networks, autonomous control of vehicles and smart grids, were not the concern of these generations.

The fifth step in the communication network revolutions has recently started. It arguably envisions a new form of proximity-aware networking or ubiquitous computing~\cite{corson2010toward,corson2013flashlinq}.
The vision is founded on the implementation of a ``wireless sense" using low-latency and highly reliable proximal communications
via device-to-device and short-range communication technologies~\cite{brahmi2015deployment},
thereby making the \ac{IoT} a technological possibility.
The ``things", including sensors and actuators, smart meters,
\ac{RFID} tags and other devices, are interconnected together either directly or through a gateway node and the global Internet. In such emerging \ac{IoT} communication systems, low latency and high reliability in communicating messages will have a large impact on all elements of \ac{ICT}, ranging from mobile networks, consumer broadband, video, wireless sense-based proximal and cloud services. Ultimately, the fifth step of the communication network revolution is expected to ensure end-to-end communication latencies below $1$ ms.

There are many use cases demanding below $1$ ms low-latency communications. The development of \ac{ICT} for healthcare, industrial processes, transport services or entertainment applications, generates new business opportunities for network operators~\cite{lema2017business}.
Part of this vision is grounded on the future availability of very low-latency communication networks to build the Tactile Internet that will extend touch and skills, and help realize real-time virtual and augmented reality experience \cite{Simsek:16}. The transmission of multi-sensorial signals, including the sense of touch (haptics), will contribute to the overall experience of real-time remote interactions. The healthcare industry is experimenting with remote diagnosis with haptic feedback,
while remote robotic surgery with haptic feedback represents a potential future application of major impact for low latency communications.
The transport sector is testing driver assistance applications and self-driving cars that will benefit from remote monitoring.
The entertainment industry expects that immersive entertainment and online gaming incorporating \ac{AR}
will open new revenue streams, while the manufacturing industry expects significant productivity increase
due to remote control with \ac{AR} applications.
Remote control applications can help improve the safety of personnel and reduce the cost of managing the on-site work force for hazardous environments such as mines or construction sites.

The use cases mentioned above, from the perspective of the wireless access network design, are expected to be addressed by the fifth step of the network revolutions, which is commonly called as \ac{5G}. \ac{5G} is driven by the vision of the networked society~\cite{Zaidi2017Designing}, for which two generic communication modes of \ac{MTC} will be supported~\cite{Torsner2015Industrial}: \ac{URLLC} and massive \ac{MTC}.
\ac{URLLC} is an innovative feature of \ac{5G} networks,
as it will be used for a range of mission critical communication scenarios. Thereby, \ac{URLLC} is expected
to create near-term new business and service opportunities.
Therefore, there is a great interest in technologies that will enable
proximal communication links as well as global networks to operate with very low-latency (below milliseconds) while ensuring high reliability.
As emphasized in~\cite{Redpath2017Monetizing}, communication networks with latency below milliseconds will have a direct impact on network and proximal communication technology monetization, because latency performance is decisive for a number of revenue generating services, including high capacity cloud services, mission critical machine type communication services and high resolution video and streaming services. Indeed, the latency performance will be the determining factor between winning and losing business since meeting latency requirements directly impacts latency-sensitive, high capacity proximal as well as global Internet services.

Unfortunately, the vision of low-latency communication networks appears as a problem of formidable complexity. Ensuring that audio, visual and haptic feeds are sent with sufficiently low latency is a challenge, since the end-to-end route may incorporate multiple wireless access, local area and core network domains. All services delivered over proximal or long-distance communication networks
are subject to latency, which is a function of several factors, such as link sharing, medium access control and networking technologies,
competing service and traffic demands, or service-processing algorithms. Within a single network, there are several components that contribute to latency, such as at the physical, link and routing layers. If we try to ensure low latency only at one layer, we may have non-negligible latency components at other layers. Moreover, the optimization at a single layer may have undesired effects at other layers. To make the problem worse, complexity is not limited to individual networks. The many use cases demanding low latency as described above, use end-to-end connections that may be supported not only over a single network, but often over multi-domain networks.
Here, ``domain" refers to a part of the end-to-end communication network that is under the control of a network operator
in terms of dimensioning and managing communication resources.
Examples of multi-domain networks include the global \ac{PSTN},
inter-networks consisting of multiple networks owned and operated by multiple Internet service providers,
and cellular networks connecting cellular subscribers served by different mobile network operators. Such a complexity is largely unexplored.

In this paper we conduct an analysis and classification of the most prominent design principles and  enabling technologies that are needed
to meet the stringent requirements of low-latency networks.
While it seems probable that such networks will be initially delivered in the proximity of communicating entities~\cite{corson2010toward,corson2013flashlinq},
we also argue that there are strong business cases to deliver mission critical services even over long distances, such as for industrial remote operations,
health care and intelligent transportation systems. Future low-latency services will be provided not only in limited geographical areas, but also over multi-domain networks. Therefore, it will be useful to analyze the latency and reliability requirements
of the most important use cases for low latency
and the sources of end-to-end latency for those cases. There is a need of substantial research, standardization and development of technology enablers that are applicable at the physical, medium access control, network and transport layers in all segments of communication networks, including the access, core and service networks. Our analysis will greatly help in laying the foundation for developing technologies that can realize networks supporting services below $1$ ms latency.
Although all of these use cases require low latency, they may pose different level of reliability requirement. e.g., the reliability requirement of decentralized environmental notification messages for vehicular communications is more relaxed compared to that of Tactile Internet. More relaxed reliability requirements may render more options of the techniques to achieve the same level of latency performance.

The remainder of this article is structured as follows.
Section~\ref{Sec:Requirements} examines the latency requirements imposed by latency-critical applications and services. Then, based on this examination, the causes of latency components in single hop, multi-hop and multi-domain networks are examined and formally defined in Section \ref{sec:definition_causes}.
Section \ref{sec: Intra-network} surveys technology enablers applicable within a single domain, while
Section \ref{sec: Inter-network} discusses inter-domain latency reduction and control techniques.
Next, Section \ref{Sec:Standards} provides an overview of the most important related
standardization activities. Section \ref{Sec:Discussion} discusses open research questions, and Section \ref{Sec:Conc} concludes the article.

\section{Latency Requirements of Use Cases in Single Hop, Multi-Hop and Multi-Domain Networks}
\label{Sec:Requirements}

In this section, we analyze the latency requirements of future use cases of major societal impact, which will be enabled by the availability of networks capable to offer below $1$ ms latencies. We argue that some of these use cases require a single domain network, whereas the rest should be supported by multi-domain networks.

The characterization of end-to-end latency and the identification of latency requirements associated with latency-critical applications and services
are necessary steps to understand and compare promising technology enablers.
We note that the definition of communication latency is not unique, but it depends on the use cases.
Due to the stochastic nature of end-to-end latency, latency requirements may be typically specified in the form of stochastic measures,
such as the cumulative distribution function or its moments~{\cite{weiner2014design,yamamoto201560}, and a probability of exceeding a predefined latency value.
For mission critical and industrial process control systems, for example, the latency requirements
can specify that a predefined latency value of a few milliseconds should be kept with probability $10^{-8}$~\cite{weiner2014design}.
Alternatively, latency requirements may require that the expected value
and the variance of the latency must remain under predefined thresholds~\cite{yamamoto201560}.
The latency requirements of prominent latency-sensitive application scenarios are listed in Table~\ref{Table:latency requirement}.
In the following, we give a short description of each.

\begin{table*}[t]
    \begin{center}
    \caption{\sc{The requirement for low-latency applications (reliability marked with high is due to no availability of precise numbers).}}
    \begin{tabular}{llll}
    \Xhline{.8pt}
    &  Latency & Reliability & Other \\
     \Xhline{.8pt}
    Virtual reality~\cite{Maier2016} & 1~ms & - &  high data rate \\
    Automated guided vehicle~\cite{Fettweis2014a,OSSEIRAN2015} & few ms & 99.99999\% & high data rate \\
    Financial market~\cite{Hasbrouck2013} & few~ms & high & \\
    Exoskeletons and Prosthetic hands~\cite{Fettweis2014a,Bogue2009,Tyler2016} & few ms & high & -\\
    Tele-surgery~\cite{Fettweis2014a,Tozal2013} & 1-10~ms& 98\% & high data rate\\
    Protection traffic in smart grid~\cite{ETRI2012} & 1-10~ms & high & - \\
    Factory automation~\cite{orfanus2013ethercat,OSSEIRAN2015}  & 1-10~ms & 99.9999999\% & - \\
    Control traffic in smart grid~\cite{ETRI2012} & 100~ms & high & - \\
    Process automation~\cite{liu2017effects,luvisotto2017high,OSSEIRAN2015} & 100~ms-1s & 99.9999999\% & - \\
    \Xhline{.8pt}
    \end{tabular}
    \vspace*{0mm}
        \label{Table:latency requirement}
    \end{center}
\end{table*}

\begin{figure*}[t!]
  \centering
  \includegraphics[width=2\columnwidth]{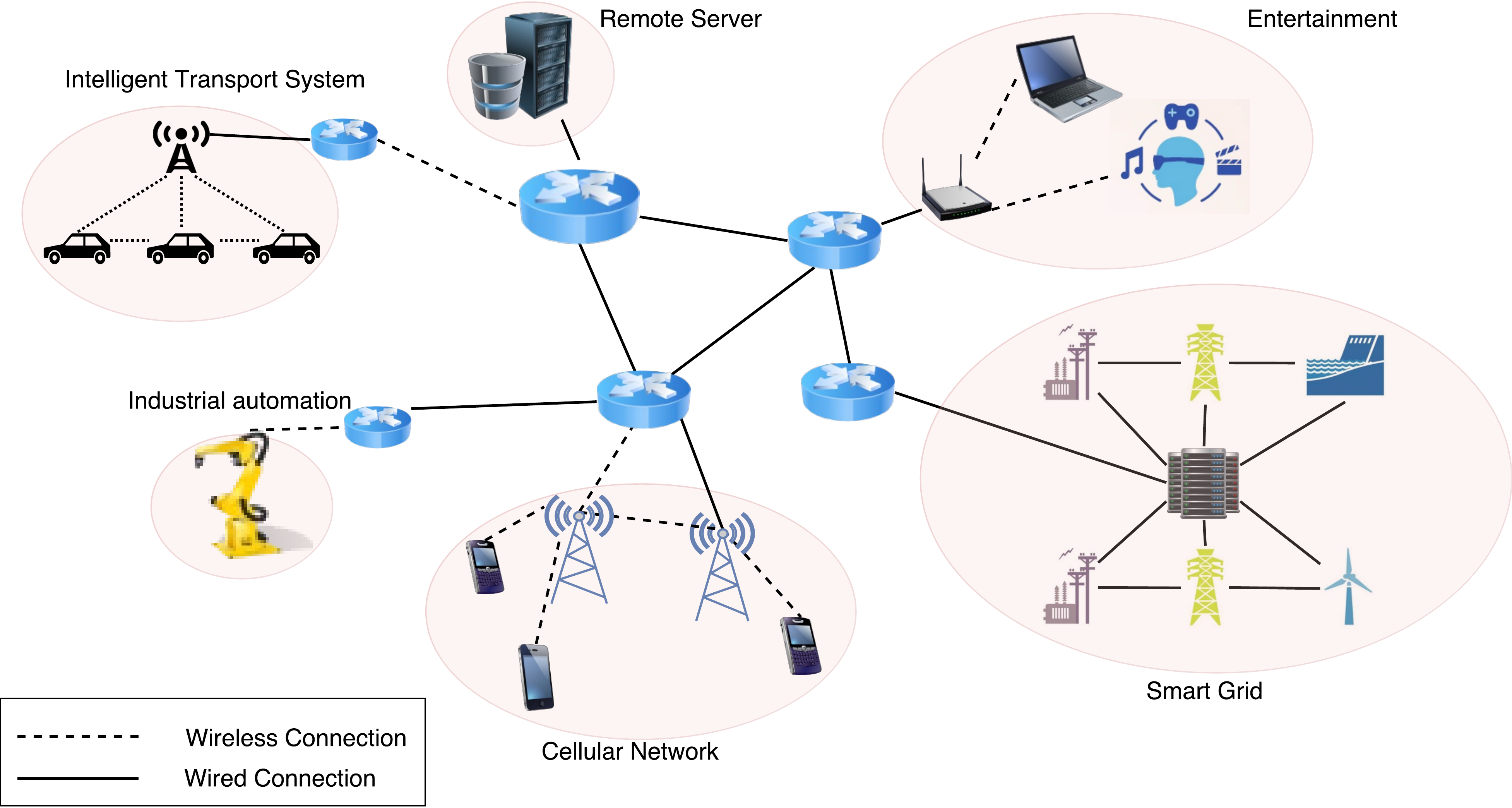}

  \caption{General network architecture. The end devices are connected to base stations or access points, which further communicate with one another through router switches mostly by wired networks.}\label{fig:networkstructure}
\end{figure*}

\subsubsection*{Smart Grid}

Smart grid is proposed to solve the deficiencies in the old power system such poor controllability to the power generation utilities
and slow response to the change of the power consumption~\cite{Guengoer2011}.
Besides the power grid, the smart grid also has a control network with monitoring, communicating and computing abilities.
This control network should support low-latency communications to enable mission critical applications such as substation automation and distributed energy resources~\cite{khan2013comprehensive}.
Specifically, substation automation refers to monitoring, protection and control functions performed on substation and feeder equipments. The substation automations are extremely latency-sensitive as the time critical message will be irrelevant if not delivered within a specific time in the order of several millisecond.
Distributed energy resources are small sources of power generation and/or storage that are located close to the load they serve and connected to the distribution grid. To integrate the distributed energy resources smoothly, automatic and remote monitoring, control, manipulation and coordination should be performed in real time.
Protection and control traffics are the most latency-sensitive in smart grid applications, with latency requirements of 10~ms and 100 ms, respectively~\cite{ETRI2012}.
IEC 61850 is the international standard that ensures interoperability between the control networks of smart grids~\cite{IEC61850}.

\subsubsection*{Industrial Automation}

The forthcoming Industry 4.0 paradigm is expected to substantially boost interoperability in manufacturing by enabling machines, devices, sensors and people to connect and communicate with one another. Industrial manufacturing requires high reliability and stringent latency guarantees.
By supervising the production activities, process automation aims to support more efficient and safe operation of industries such as paper, mining and cement~\cite{liu2017effects,luvisotto2017high}. The latency requirements are related with the sampling times of different applications and are in the order of 100 ms to 1 s.
Factory automation includes time-constrained operational applications, such as those used for motion control and certain power electronics applications, which requires the latency between 1-10 ms~\cite{orfanus2013ethercat}.
Ethernet-based solutions are gaining popularity in industrial communications, due to their capability of guaranteeing latency.~\cite{wollschlaeger2017future}.
The wireless solutions are also favored in many industrial scenarios with harsh environments.

\subsubsection*{Medical Applications}
Tele-diagnosis and tele-surgery are promising trends for medical care. Not constrained by the geographical distances anymore, experienced surgeons will be able to diagnose or even perform surgeries along with audio-visual and haptic feedback by a robot~\cite{Fettweis2014a}.
Tele-monitoring is another emerging application that enables the experienced surgeon at the remote side to watch both the local surgeon to perform the surgery, and the patient's conditions, and provide real-time guidance and suggestions.
These use cases require, in general, latencies of some  milliseconds and at most 2\% packet loss rates to realize real-time and reliable audio, visual, and haptic feedback~\cite{Tozal2013}.

Medical applications of low-latency networks go way beyond tele-surgery. The exoskeletons are new supportive protheses that enable to help aging people move independently, promote patient rehabilitation following an injury, and allow workers to carry heavier loads. Exoskeletons use sensors on the skin to detect voltage change of the signal at the muscles sent by the brain~\cite{Bogue2009}.
Prosthetic hand is another artificial device, designed for those that lost their own hand. Most of the existing prosthetic hands can only perform actions such as bracing and holding, which are control actions that can be performed with some hundreds of milliseconds' communication latency between the touching and the actuation. A prosthetic hand that enables the users to feel has been recently reported in~\cite{Tyler2016}. The tactile and position sensors in such a prosthetic hand collect and send the data to a control unit that translates them into a neural code, which is then applied to the nerves in the user's arm. Meanwhile, the data is also processed to interpret the user's movement intention and send commands to the prosthetic hand.
To provide the user with safety and a better experience wearing the exoskeletons and prosthetic hands, the latency should be no more than a few milliseconds.

\subsubsection*{Virtual Reality and Augmented Reality}
\ac{VR} and \ac{AR} are revolutionary interfaces that provide unprecedented experiences and enable new applications.
\ac{VR} provides an immersive experience for a live concert, sports match or interactive game for users just sitting on the couch at home.
\ac{AR} augments reality by enabling better learning and working modes. For example, in a natural history museum, on top of the specimen of a dinosaur, some vivid three dimensional dinosaur projection provides a better conceptual learning environment.
The real-time audio, visual, and haptic feedback by \ac{VR} and \ac{AR} requires latency from action to reaction below 1~ms to avoid nausea~\cite{Maier2016}.
Within the world of musical instruments and music industry, the vision of Internet of Musical Instruments has recently been proposed. According to such a vision, any musical instrument will be connected in the future to Internet via wireless communications, provided that the end-to-end latency is around 5~ms~\cite{turchet2016smart,turchet2017examples,turchet2017towards}.

\subsubsection*{Intelligent Transport Systems}

Vehicular communications refer to the communication among vehicles that can improve driving safety, reduce traffic congestion and traffic accidents, improve fuel consumption efficiency (e.g., by platooning), support high quality entertainment, and ultimately enable driver-less cars~\cite{santa2009sharing,alsabaan2013optimization,yu2013toward,gerla2014internet}.
\ac{AR} can also be used in intelligent transport systems to create a bird view of the real-time traffic information.
Vehicular cloud architectures are also proposed to support applications that require large computation and storage capacities~\cite{yu2013toward}.
To support all these functionalities, communication networks have to support latencies of only a few milliseconds~\cite{Fettweis2014a}. To ensure safety, a reliability as high as 99.99999\% is also needed~\cite{OSSEIRAN2015}.
Dedicated Short Range Communications based on IEEE 802.11p~\cite{Uzcategui2009} and cellular vehicle-to-everything (C-V2X) communication are promising candidates to support vehicle-to-vehicle and vehicle-to-roadside communication~\cite{5Gwhitepapervehicle,festag2015standards}.

\subsubsection*{Financial Markets}

Time equals money in the financial market, and having low-latency between the placing of an order and its execution is essential to achieve the transaction at the desired price before the price changes. It has been stated that a 1-millisecond advantage in trading applications can be worth \$100 million a year to a major brokerage firm~\cite{martin2007wall}.
The required latency in the financial trading environment is in the order of several milliseconds~\cite{Hasbrouck2013}. A high availability is also required to support a great number of online clients.

We conclude this section by noting that the major use cases mentioned above do not have a common definition of the latency requirement. Moreover, in some instances they require proximal communications, and in other instances they require remote communication services. In the next section, we will deepen the technical definition of latency and of delay components for proximal and remote communication services.

\section{Causes and Definitions of Latency Components in Single Hop, Multi-Hop and Multi-Domain Networks} \label{sec:definition_causes}

In the previous section we have seen that the use cases of major societal impact enabled by low-latency networks will have both proximal and remote communication sessions.
As latency is a complex function of several factors, it is useful to examine the components of end-to-end latency in proximal communication scenarios,
as well as over long distances involving wired and wireless access networks and multi-domain inter-networks, as illustrated by Fig.~\ref{fig:networkstructure}, which we explain shortly below. We will often refer to this figure in the rest of the paper, with the purpose of analyzing which design principle and which enabling technology are responsible for the various latency components.

In the generic multi-domain scenario depicted in Fig.~\ref{fig:networkstructure}, information messages are transported by several networks employing diverse technologies.
For example, when streaming an online video on a cellular phone, the corresponding service request may first be sent
through a single-hop wireless connection to the serving \ac{BS}.
Subsequently, it may go through a series of wired/wireless connections in the backhaul network and a set of intermediate networks to an application server.
As another example, communications among hundreds of sensors
and actuators inside a vehicle and between vehicles and vulnerable road users
and road-side infrastructure equipment are inherently local and mission critical.
Although both examples impose latency requirements, it is clear
that the requirements imposed by mission critical services can be order(s)
of magnitude lower than quasi real-time entertainment services.

We can formally define latency as the time duration between the generation of a packet
and its correct reception at the destination. This message may go through several networks as shown in Fig. 1,
sometimes referred to as domains, which may not be handled by the same operator.
It is difficult to guarantee a predefined total latency, as different networks may
introduce random latencies whose exact values or even statistical distributions are typically not known a priori.
In other words, even though the latency can be measured or estimated within a single network, meeting end-to-end latency guarantees remains a challenging task, since no single network operator or network entity has control over the end-to-end latency.

\begin{figure}[t]
\centering
\includegraphics[width=0.85\columnwidth]{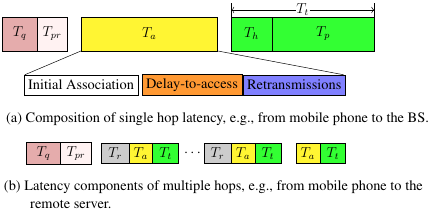}
\caption{The latency formation of single hop and multi-hop communication.}
\label{fig:delayformation}
\end{figure}

To gain a precise insight, consider Fig.~\ref{fig:delayformation}, which illustrates the most important components of latency.
Once a packet is generated, it is typically placed in a queue, and waits for transmission.
Analogously, when a packet arrives at the receiver, it may be placed in a queue at a low layer of the protocol stack,
waiting to be processed and delivered to higher layers.
We define the sum of these queuing latency as $T_q$.
To facilitate link sharing between multiple traffic classes, modern access and core networks provide differential treatments of packets,
and implement \ac{QoS} dependent packet handling, including, for example, \ac{QoS}-aware scheduling and
priority queueing mechanisms.
When multiple \ac{QoS} classes with different latency priorities are supported, the packets with higher priority have lower $T_q$.
Similarly, the total processing time during the end-to-end communication can be conveniently characterized by the aggregate processing latency  $T_{\mathrm{pr}}$.
The processing latency $T_{\mathrm{pr}}$, different from $T_q$,
is a function of physical, link layer and hardware technologies, node processing capacities and signal processing algorithms.

Apart from $T_q$ and $T_{\mathrm{pr}}$, for a single hop, latency incorporates the time it takes for a packet
to get access to the –- typically shared -– medium (denoted by $T_a$),
including the time taken by technology-dependent control signaling. For example,
successfully transmitting and receiving request-to-send and clear-to-send messages of the popular IEEE 802.11 protocol family
or processing scheduling grant messages of the 3GPP \ac{LTE} protocol suite contribute to the latency over a single hop
of an end-to-end path.
Once the (wired or wireless) medium is accessed for the delivery of a packet,
we need to account for the transmission of that packet,
which typically includes a packet header and payload, amounting to a transmission time of $T_t = T_h + T_p$.

When the end-to-end path involves multiple hops -- each of which is carried over dedicated or shared resources --
the end-to-end latency is determined by the sum of the associated medium access ($T_a$) and packet transmission ($T_t$)
times.
In addition, for multihop communications, the time required for routing packets to the right outgoing interface ($T_r$)
adds to the end-to-end latency, as illustrated in Fig.~\ref{fig:delayformation}.
The routing ($T_r$) latency may be zero when the packet is forwarded by an entity that does not perform routing,
which is the case of a wireless relay or other entity that operates at the \ac{MAC} layer.

Recall that the main sources of latency variations
-- in addition to unpredictable
interference levels, random appearance of shadowing objects, and other factors appearing on the wireless interface --
include the random traffic load along shared
links and media as well as the variations of the fast fading wireless channels.
These random factors
make the end-to-end latency notoriously difficult to control and predict
since the transmission of a tagged data stream is affected
by the fluctuating load pattern of simultaneously delivered traffic streams~\cite{massoulie1999bandwidth}.
In the following two sections, different techniques for intra networks and inter networks as shown in Fig.~\ref{fig: structure} will be investigated.

\begin{figure*}[t]
\centering
\includegraphics[scale=0.6]{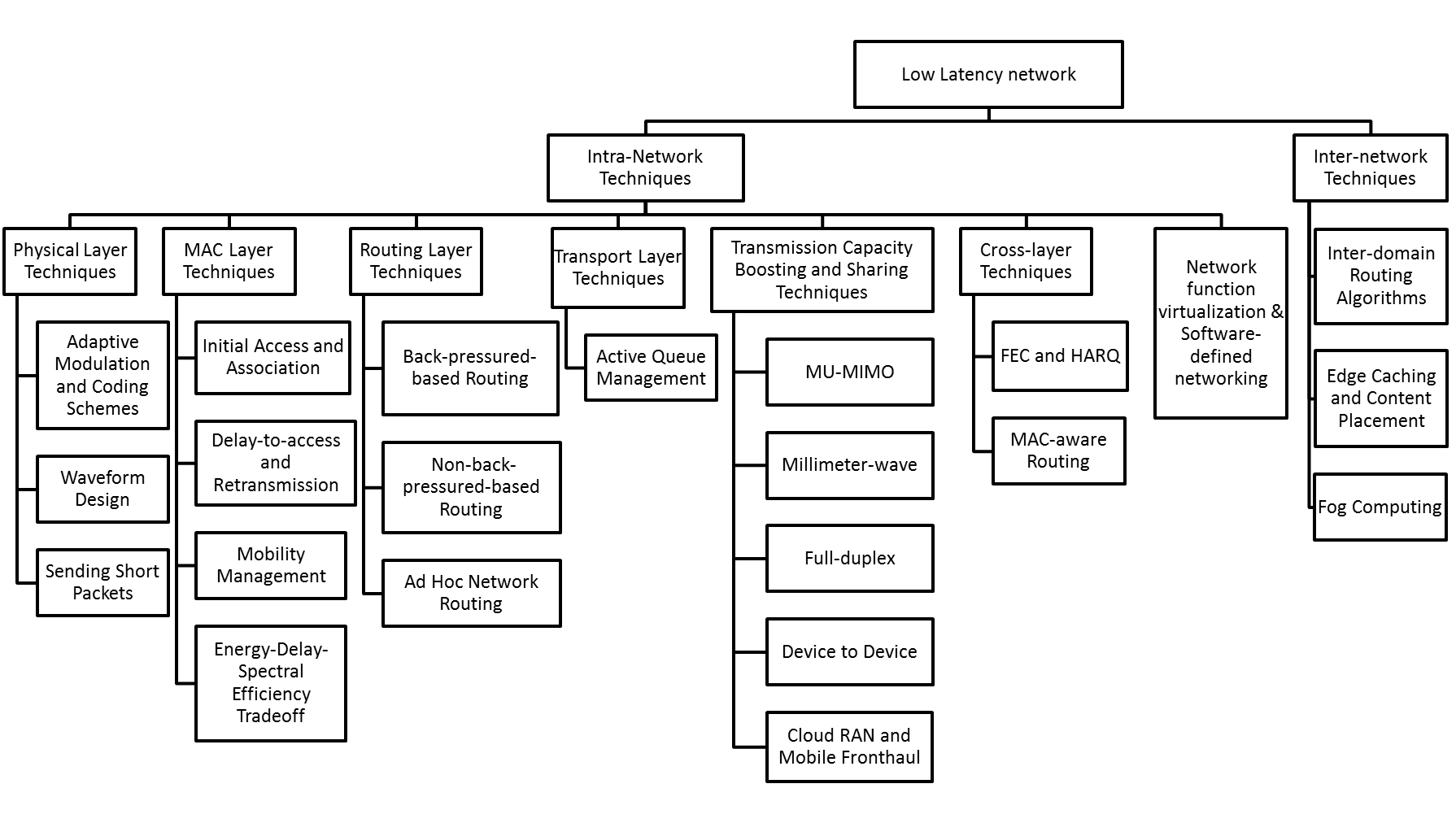}
\caption{Enabling technologies for low-latency communication, each of them may affect single or multiple delay components.}
\label{fig: structure}
\end{figure*}

\section{Intra-network Techniques and Technologies}\label{sec: Intra-network}

As illustrated in Fig.~\ref{fig:delayformation}, the end-to-end latency is an aggregate of the latency components that are associated with queueing and processing ($T_q+T_{\mathrm{pr}}$), medium access ($T_a$), transmission ($T_t=T_h+T_p$) and routing ($T_r$). In low-latency networks, techniques at the physical, medium access control, network and transport layers will have to be designed to reduce these components (see Table~\ref{Table:intra}), also in a cross-layer collaboration. These techniques will have to be tailored to reduce a specific component or deal with a combination of them.

In this section, we discuss physical, medium access control, network and transport layer techniques that can be used to reduce latency, and at the same time we take into consideration constraints related to spectral efficiency, energy efficiency, peak data rate and capacity. Also, this section discusses technology enablers that are not specifically designed to deliver latency-critical services, but can be potentially used to meet latency targets.

\begin{table*}[t]
\centering
\caption{\sc{Potential methods to achieve low-latency in a single network.}}
\label{Table:intra}
\renewcommand\arraystretch{1.2}
\begin{tabular}{lccccc}
\Xhline{.8pt}
 Methods                   & Transmission time      & Medium access latency      & Routing latency     & Queueing latency      & Processing latency  \\
                           &$T_t$                   &$T_a$                       &$T_r$                &$T_q$ &$T_{\mathrm{pr}}$ \\
 \Xhline{.8pt}
 Adaptive modulation $\&$ coding     & $\surd$                  &                                   &                               &                              &                       \\ \hline
 Waveform design                     & $\surd$                  &                                   &                               &                              &     $\surd$           \\ \hline
 Sending short packets               & $\surd$                  &                                   &                               &                              &                       \\ \hline
 MAC algorithms                      &                          & $\surd$                           &                               &                              &                       \\ \hline
 Routing algorithms                  &                          &                                   & $\surd$                       &                              &                       \\ \hline
 Transport Layer algorithms          &                          &                                   &                               & $\surd$                      &                       \\ \hline
 Capacity boosting $\&$ sharing                    & $\surd$                  & $\surd$                           &                               &                              &                       \\  \hline
 Device to device communication      & $\surd$                  & $\surd$                           &                               &                              &                       \\ \hline
 Cloud RAN $\&$ Mobile Fronthaul     &                          &                                   & $\surd$                       &                              & $\surd$               \\ \hline
 FEC $\&$ HARQ                       & $\surd$                  & $\surd$                           &                               &                              &                       \\ \hline
 MAC-aware Routing Algorithms        &                          & $\surd$                           & $\surd$                       &                              &                       \\ \hline
 NFV $\&$ SDN                        & $\surd$                  & $\surd$                           &  $\surd$                     &$\surd$                       &$\surd$                  \\
\Xhline{.8pt}
\end{tabular}
\end{table*}

\subsection{Physical Layer Techniques}

Physical layer (PHY) techniques are fundamental in striking a good engineering trade-off between latency, reliability, spectral and energy efficiency, and communication range. To quantify the effects of PHY techniques on latency, recall that, assuming a fixed-length packet, an higher transmission rate implies lower transmission time ($T_t$). In a single antenna system the maximum transmission rate between a transmitter and its intended receiver is upper bounded by
\begin{equation}\label{Shannon}
  C = B \log_2(1+\mathrm{SINR})\hspace{2mm} [\text{bps}] \:,
\end{equation}
where $B$ is the bandwidth, and the instantaneous \ac{SINR} is
\begin{equation}\label{SNR}
  \mathrm{SINR} =\frac{|h|^2 P_t}{N_0 B + P_I} \:,
\end{equation}
where $h$ is the instantaneous channel gain, $P_t$ is the transmit power, $N_0$ denotes the noise spectral density, and $P_I$ is the received interference power.

Adaptive modulation and coding is a powerful technique to increase the spectral efficiency or decrease the \ac{BER}. Generally speaking, for sufficiently high \ac{SINR}, higher modulation orders together with light coding schemes boosts the transmission rate and reduces $T_t$. However, when the channel becomes poor or the received interference is high, the transmitter should reduce the transmission rate (by adopting lower modulation orders or stronger coding schemes) to maintain a target \ac{BER}~\cite{Yilmaz2015}.

From (\ref{Shannon}) it is straightforward to see that the  bandwidth ($B$) acts as a multiplication factor in the transmission rate calculation, thus increasing the bandwidth is an effective way to increase the transmission rate and thereby to reduce latency. As $B$ also affects $\mathrm{SINR}$ from (\ref{SNR}),
the capacity is not a linear function of the bandwidth. However, we focus on the range where capacity increment with bandwidth is not negligible, i.e., it has not yet saturated. Despite that this is the range in which the reliability is not very high, there could be an interesting trade-off among capacity, latency, and reliability.
Taking \ac{mmWave} communication as an example, the abundant bandwidth at \ac{mmWave} frequency bands enables it to achieve multi-gigabit transmission rates -- even with a very low order of modulation -- at the expense of lowering spectral efficiency.

\subsubsection{Adaptive Modulation and Coding Schemes}

The randomness of the fading wireless channel, $|h|$ in (\ref{SNR}), causes the SINR at the receiver to fluctuate. However, when the transmitter is able to acquire \ac{CSI}, it can adapt its transmission parameters, which helps to achieve high data rates, which in turn helps to reduce $T_t$ by adaptively setting the transmit power, modulation scheme, coding rate, or the combinations of these parameters. Indeed, as it was shown in~\cite{Goldsmith1997}, adaptive transmission schemes provide higher average transmission rates compared to  non-adaptive transmission schemes. This is because in order for non-adaptive schemes to operate with acceptable \ac{BER}, they need to be designed for the worst case channel conditions, and therefore operate with low spectral efficiency even when the channel condition is favorable. The seminal work in~\cite{Goldsmith1997} presented optimal and suboptimal adaptation policies, including the total channel inversion scheme that adjusts the transmission power to maintain a constant received power,
and truncated channel inversion scheme that only compensates for fading above a certain cutoff fade depth (see~\cite{Goldsmith1997a} for details).
The \ac{CSI} feedback of adaptive modulation and coding takes resources and increases the latency. However, the overhead by the \ac{CSI} feedback may be compensated by the gain of the average transmission rate. For bi-directional communication with a packet exchange period shorter than the required \ac{CSI} feedback interval, the \ac{CSI} feedback can always be piggybacked in the packets sent back to the transmitter, further reducing the \ac{CSI} overhead.

\subsubsection{Waveform Design}

Once a message (and the corresponding data packet) is mapped onto symbols with an appropriate modulation and coding scheme, selecting a proper waveform modulation scheme has a great impact on the transmission time. Indeed, recognizing the importance of waveform selection, the research and standardization community spent a large effort on analyzing the advantages and disadvantages of waveform candidates for the next generation of wireless systems~\cite{Wunder2014}.

\ac{SC} and \ac{MC} modulations are two categories of waveform modulations. With \ac{SC}, each signal is spread over the whole available bandwidth. The  transmission time for each signal is short, but without time guard protection, the delay spread in wireless channels may cause \ac{ISI} to subsequent symbols.
To combat \ac{ISI}, \ac{MC} modulation divides the available bandwidth into multiple narrow sub-carriers, and maps one modulated signal to each sub-carrier. The transmission time of each signal increases, making the delay spread relatively short and presenting a higher \ac{SINR} at the receiver side.
However, the overall transmission time of the payload of \ac{MC} is comparable to that of \ac{SC}.

\ac{OFDM} is a widely used \ac{MC} modulation technique employed by IEEE and 3GPP systems. It employs orthogonal sub-carriers to achieve bandwidth efficiency. A brief comparison between OFDM and SC in terms of latency is given below.
First, the transmission rate of each sub-carrier in \ac{OFDM} is smaller than that provided by \ac{SC} modulation. However, the aggregate transmission rate over all sub-carriers is comparable to that of \ac{SC}. In OFDM a \ac{CP} is used to increase robustness against \ac{ISI} in multipath propagation environments.
Although the overhead introduced by the \ac{CP}, which must be greater than the maximum delay spread of the channel, is not negligible, it is still  more economical than \ac{SC} modulation where one guard time interval is needed for each symbol~\cite{banelli2014modulation}.
Apart from the high achievable spectral efficiency in multipath propagation environments, \ac{OFDM} has the flexibility of adaptively assigning different transmission powers and symbol constellations to each sub-carrier, which helps strike a good balance between transmission rate and reliability in frequency selective environments. In terms of processing latency, \ac{OFDM} leads to somewhat longer latency than block processing at both the transmitter and the receiver, but its equalization can be done efficiently on a sub-carrier basis using a simple one-tap equalizer. A more detailed analysis can be found in~\cite{pancaldi2008single} and the references therein.

The orthogonality of the sub-carriers is essential for \ac{OFDM} to avoid inter-carrier interference, which requires high precision synchronization between the transmitter and the receiver. Thus traditional \ac{OFDM} with long synchronization is not ideal for meeting stringent latency requirements in a spectral efficient manner when transmitting short packets. The recently proposed \ac{WirelessHP} aims to support latency in the order of $\mu$s for industrial control applications using short packets~\cite{luvisotto2017physical}. \ac{WirelessHP} reduces the physical layer header of OFDM by taking advantage of the predictive and periodic traffic pattern of industrial applications.

The research community has developed new waveforms that help relax the strict synchronization requirement, e.g., \ac{UFMC}, \ac{FBMC}, \ac{GFDM}~\cite{Wunder2014}, and \ac{F-OFDM}~\cite{Abdoli2015}. These waveforms are non-orthogonal and thus inter-carrier interference must be kept under control by using filters to suppress the out-of-band emission.
Among the above asynchronous waveforms, \ac{GFDM} is regarded as the most suitable for low-latency communication~\cite{Farhang2014}, \cite{Fettweis2009}. Compared with \ac{OFDM}'s division only in the frequency dimension, \ac{GFDM} has a block frame structure composed by $M$ sub-symbols and $K$ sub-carriers, and each sub-carrier is modulated and filtered individually.

For low-latency communications, \ac{GFDM} acts as a comparable candidate to \ac{OFDM} for several reasons.
First, using \ac{GFDM} may achieve lower $T_h$, as it requires less synchronization accuracy compared to OFDM.
Second, the duration of the \ac{GFDM} symbol is more flexible than in OFDM, and may achieve lower $T_p$.
To combat \ac{ISI}, a \ac{CP} is appended to each \ac{OFDM} symbol, while non-orthogonal \ac{GFDM} needs a single \ac{CP} to protect the information contained in $M$ sub-symbols. To compare the $T_p'$s of \ac{OFDM} and \ac{GFDM}, set the number of sub-carriers of \ac{OFDM} $N$ as the product of the numbers of sub-carriers and of sub-symbols ($N = M \cdot K$), which suggests that the sub-carrier spacing of \ac{GFDM} is $M$ times that of \ac{OFDM}, and the sub-symbol duration of \ac{GFDM} is $1/M$ that of \ac{OFDM}. Given a \ac{CP} duration, when mapping $N$ bits of data, the $T_p'$s of \ac{OFDM} and \ac{GFDM} are the same. However, if $N+1$ bits of data are sent, \ac{GFDM} can easily be adapted to $(M+1)$ sub-symbols while still using a single \ac{CP}. On the other hand, \ac{OFDM} can change the sub-carrier size to $K$ resulting in $(M+1)$ \ac{OFDM} symbols, each having one \ac{CP}. This is equivalent to $M$ extra \ac{CP} durations compared to \ac{GFDM}.
Third, distortion accumulation can grow without bounds in OFDM when orthogonality is not perfectly maintained, so OFDM must have proportional sub-carrier spacing to guarantee orthogonality~\cite{Wunder2014}. \ac{GFDM} inherently deals with non-orthogonality, and can therefore use non-proportional sub-carrier spacing, and non-continuous spectrum aggregation. Thus, \ac{GFDM} can also be used to improve the transmission rate.
On the negative side,
\ac{GFDM} requires high transmitter filter order to achieve sharp filter edge, which increases both complexity and processing latency.
To summarize, the ideal waveform should be determined by the design criterion and operational environment, including the trade-off between reducing the transmission time and increasing complexity and processing latency.

\subsubsection{Sending Short Packets}
\label{Subsubsection:short-packets}
It is intuitive to think that short packets are needed for short transmission times in latency sensitive communications. The use of short packets brings a difference to the maximum coding rate and the packet error probability~\cite{Durisi2015}.
Specifically, for a given packet error probability and for a finite packet length $n$, the maximum coding rate reduces by a factor proportional to $1/\sqrt{n}$~\cite{Polyanskiy2010a}. The maximum coding rate for finite packet length with multiple antennas when considering the trade-off of diversity, multiplexing, and channel estimation is investigated in~\cite{Ostman2014a, Durisi2014a}.
The maximum coding rate affects the minimum packet transmission time $T_t$, and further affects the end-to-end latency.

With traditional long packets, the payload data is much longer than the metadata (control information), and thus the metadata is coded with a low rate to be robust. As the meta data accounts for only a small fraction of the whole packet, the overhead caused by the low coding rate can be neglected since it virtually does not affect the latency performance.
With short packets, however, the size of meta data $T_h$ is comparable to the size of payload data $T_p$.
In this case, the overhead introduced by the meta data cannot be neglected, rendering coding metadata with a low rate inefficient. Therefore, to increase the spectral efficiency, metadata and payload should be coded jointly, the details of which is an open research issue~\cite{Durisi2015}.

\subsection{MAC Layer Techniques}
\label{Subsection:MAC}

The \ac{MAC} layer is responsible for synchronization, initial access, interference management, scheduling, and rate adaptation. While~\eqref{Shannon} governs the maximum achievable rate, inefficient initial access, queue management, and channel access strategies may substantially reduce the effective transmission rate of individual devices.

Define the one-hop access latency for every packet as the time from the instant the node starts sending that packet for the first time until the beginning of its successful transmission. It includes the processing and queuing latencies, which we do not cover in this paper, the association latency (only in the initial access phase), delay-to-access (latency until the start of the scheduled time slot in contention-free manner or that before the first transmission attempt in the contention-based manner), and the retransmission delay in the case of decoding failure. In the following, we review main techniques used to reduce these latency components. As we will see, this reduction comes (usually) at the price of a higher computational and signaling overheads, and also a penalty in energy and spectral efficiencies.

\subsubsection{Initial Access and Association}
Initial access and association are amongst the most important \ac{MAC} layers functions that specify how a new device should connect to the network. This is usually handled by a synchronization process and then a random access phase, by which the network registers the device as active.
The latency caused by the association procedure may be tolerable when the devices have to be connected all the time (e.g., mobile phones), the data size is large (e.g., camera sensor networks), or the handover time is negligible; see also Section~\ref{sec: handover}. In many \ac{IoT} applications with massive number of wake-up radios each having just a few bits of data, however, the association latency may become much longer than the data transmission time.
Unfortunately, most of the currently used standards are not capable of supporting a low-latency initial access procedure: the initial access deadline of 10~ms for 3GPP~\cite{soldani20185g} or 20~ms for ITU~\cite{parkvall2017nr}. That is why the existing standards only consider ``connected users'' for low-latency services and assume ``normal'' initial access.
Designing more efficient synchronization and initial access procedures for low-latency networks seems to be widely open areas.

\subsubsection{Delay-to-access and Retransmission}\label{sec: Delay-to-access}
The \ac{MAC} layer scheduling is responsible for the delay-to-access component. A \ac{MAC} protocol is contention-free if messages do not collide during its execution, which is usually guaranteed by orthogonal communication resource allocation to different devices. The orthogonality of contention-free can be realized in the time domain (e.g., \ac{TDMA}), in the frequency domain (e.g., \ac{FDMA}), in the code domain (e.g., \ac{CDMA}), in the spatial domain (e.g., \ac{MU-MIMO}) or any combination of those domains. In contention-based \ac{MAC} protocols, devices contend to access the channel, and as a result, some messages are lost due to inevitable strong interference (also called collision).

Contention-free \ac{MAC} protocols can guarantee certain latency and jitter at the price of signaling and computational overheads, which may not be tolerable or efficient in the use cases of low-latency networks
with a large number of devices and each just has a few bits of data.
Consider the example of massive \ac{IoT} devices with wake-up radios. After being associated to the network, the radios should send a channel access request to a local coordinator (e.g., AP or BS), wait for the channel access notification and follow the instruction to transmit their few bits of information. For the \ac{TDMA} strategy applied to $N$ devices, the average latency is $N/2 \times T_t$, where $T_t$ is the time slot duration for each device. Clearly, this delay is not scalable with the number of devices. \ac{FDMA}, \ac{CDMA}, and \ac{MU-MIMO} do not have this problem but the devices should still register their channel access requests and wait for the instruction (carrier frequencies in \ac{FDMA}, codes in \ac{CDMA}, and beamforming vectors in \ac{MU-MIMO}), which could be problematic in massive wireless access scenarios~\cite{gu2009spatiotemporal}; see also Section~\ref{sec: capacity-boosting}.

In contention-based \ac{MAC} protocols, there is no controller that governs the scheduling. This class of protocols have low signaling and computational overheads, but they impose a random latency for channel establishment. Slotted-ALOHA is a very simple contention-based scheduling, in which a device tries to access the channel at the start of the next slot as soon as it gets a new packet~\cite{roberts1975aloha}. In the case of collision, it retransmits the packet after a random backoff whose average increases after every new collision event. It has a good delay and throughput performance when the offered load (traffic arrived per unit time) is low. The devices do not need to have any signaling to reserve the channel, unlike the contention-free protocols. When the offered load increases, however, the delay and throughput performance of slotted-ALOHA degrades dramatically due to continuous backoff caused by repeated collisions.

\ac{CSMA} protocol introduces channel assessment that allows each device to check whether the channel is idle before transmission. This technique, together with the random backoff procedure, substantially reduces the collision probability, leading to a better performance in high-load scenarios.
Yang {\textit{et. al.}}~\cite{Yang2003} studied the delay distribution of slotted-ALOHA and \ac{CSMA} and showed good delay performance of both schemes in low traffic regimes. \ac{CSMA} is known to have hidden and exposed node problems, which may cause collision and unnecessary deferred transmission~\cite{sohraby2007wireless}. \ac{CSMA/CA} uses \ac{RTS} and \ac{CTS} signals to reserve the channel in a distributed fashion so as to reduce collision probability.
However, the extra signaling may reduce the throughput by as much as 30\%~\cite{Magistretti2014} and increase the delay. This inefficiency becomes more prominent when we go to high data rate technologies like \ac{mmWave} networks where the transmission rate of data signals are 100--1000x higher than that of control signals. In that case, the overhead of the collision avoidance mechanisms can drop the link performance to only 10\%~\cite{Shokri-Ghadikolaei2015}.
Fig.~\ref{fig:TDMAALOHA-subfig-a} shows the latency (due to delay-to-access and retransmission) performance of slotted-ALOHA, \ac{CSMA}, and \ac{TDMA} for a network of single antenna devices containing 100 transmitters. In all scenarios, increasing the input traffics increases both network throughput and average delay, almost linearly. Once the network operates close to its capacity, increasing the input traffic would not improve the network throughput but substantially increases the delay. In contention-based algorithms there is a critical value of the input traffic after which the network becomes congested, and adding more traffic decreases the network throughput and increases the delay exponentially. From this figure, slotted-ALOHA and CSMA are preferable for light traffics, \ac{CSMA/CA} (not shown in this figure) for medium traffics, and \ac{TDMA} for intensive traffics. Note that the overhead of TDMA channel access request is not shown in this figure.

\begin{figure}[t]
  \begin{subfigure}[t]{\columnwidth}
    \centering
    \footnotesize{\input{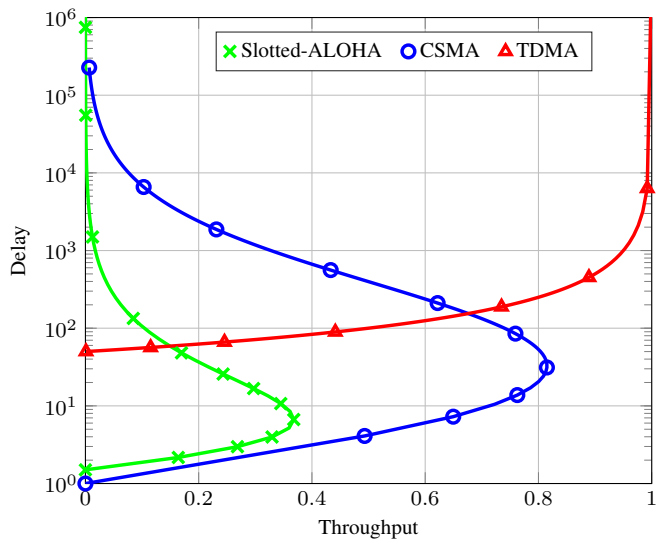}}
    \caption{Omnidirctional antenna}
    \label{fig:TDMAALOHA-subfig-a}
  \end{subfigure}
  \begin{subfigure}[t]{\columnwidth}
    \centering
    \footnotesize{\input{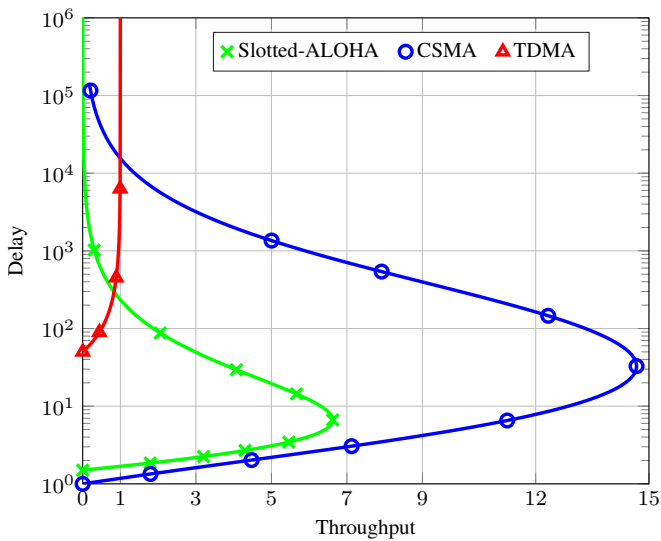}}
    \caption{Directional antenna}
    \label{fig:TDMAALOHA-subfig-b}
  \end{subfigure}

 \caption{Delay versus throughput of slotted-ALOHA, CSMA, and TDMA. Two numbers of users, 10 and 100, are set for TDMA. The unit of delay is the packet transmission time. \subref{fig:TDMAALOHA-subfig-b} is based on~\cite{Shokri2015Transitional}/}
 \label{fig:Throughput-delay}
\end{figure}

When we deal with devices with multiple antennas, the traditional trade-offs and design constraints may change dramatically~\cite{Shokri2015Transitional,Ramanathan2005Adhoc,son2012frame,Singh2011Interference,Stahlbuhk2016Topology}. Directional communication alleviates the hidden and exposed node problems and reduces the interference footprint. The collision avoidance mechanism are less important and \ac{CSMA} and slotted-ALOHA substantially outperform \ac{TDMA} in terms of throughput and delay~\cite{Shokri2015Transitional}.
The analysis in~\cite{Shokri2015Transitional} shows that directional communication is key to support low latency in massive wireless access scenarios. In particular, higher directionality levels reduce the need for complicated scheduling approaches with high signaling and computational overheads. Fig.~\ref{fig:TDMAALOHA-subfig-b} shows the delay performance of \ac{TDMA}, slotted-ALOHA, and \ac{CSMA} for a network of multiple antenna devices. Now, slotted-ALOHA and CSMA are preferable for almost any traffic, given that obtaining beamforming vectors is not time consuming. \etal{Nitsche}~\cite{nitsche2015steering} uses dual radio wherein one radio is responsible for obtaining directional information all the time to be fed to the other radio that is responsible for the data transmission. This technique decreases the delay to find proper beamforming vectors.
\etal{Olfat}~\cite{Olfat2017Learning} relaxes the dual-radio requirement and develops a reinforcement-learning approach to design the optimal beamforming vectors with a minimal number of pilots. However, the research area of low-overhead beamforming is widely open and further research is required.

An advanced technique to utilize correlatable symbol sequences in place of the traditional control messages is proposed in 802.11ec~\cite{Magistretti2014}.
Correlatable symbol sequences are predefined
pseudo-noise binary codewords that retain the statistical properties of a sampled white noise, the cross correlation of which obtains spike value only with a matching copy.
These sequences can be correctly decoded at very high transmission rates due to the robustness of the receiver. \cite{Shokri-Ghadikolaei2015} extends the use of these sequences to the \ac{mmWave} networks to address the imbalance transmission rates of the control and data planes, and improves the efficiency of distributed channel reservation for ad~hoc \ac{mmWave} networks. Interestingly, the extended approach can substantially alleviate the need for retransmission of collided channel reservation packets as different packets (of different types like \ac{RTS}-\ac{CTS} or of the same type transmitted by different devices) can be recognized at the receiver even when they arrive at the same time. This technique proves to be suitable for low-latency communications in ad~hoc networks.
Grant-free communication schemes for short packets are a new approach that is proposed in~\cite{azari2017grant,clazzer2016exploiting}. These schemes tolerate collisions by sending replica packets at the transmitter side and applying successive interference cancellation at the receiver side.
In \cite{Swamy2015}, the authors propose a frame-based contention-free protocol for star networks that exploits the cooperative multi-user diversity. Specifically, in each frame, the devices that have successfully sent their data, help the remaining unsuccessful devices by relaying their data toward the \ac{AP}. This multiuser diversity approach is also combined with network coding to further improve the reliability and delay performance~\cite{Swamy}.

\subsubsection{Mobility Management}\label{sec: handover}
Mobility management and handover are essential procedures of a mobile network to maintain a connection. During the handover, the mobile terminal switches the set of its serving \acp{AP}/\acp{BS} based on some performance measures (e.g., received signal strength indicator). When a terminal is capable of communicating to more than one \ac{AP} at a time, soft handover eliminates the need for breaking radio links. For other technologies where no such capability is supported, the mobile terminal can be served in parallel by up to one \ac{AP} and therefore has to break its communication with its current \ac{AP} before establishing a connection with a new one. Traditionally, reducing the signaling overhead and algorithmic complexity of handover were the primary design concerns of almost all existing commercial systems.

References~\cite{choi2005fast,lee2006fast,jiao2007fast} proposed various algorithms for fast hard handover algorithms in cellular networks. The general idea of all these schemes is to anticipate handover events and perform some operations prior to the break of the radio link. More precisely, these references focused on mobile broadband wireless access (in particular IEEE~802.16e) and introduced different mobility management messages that enable receiving data packets during the handover process. \etal{Li}~\cite{li2008enhanced} generalized the idea of fast handover to mobile IPv6 to maintain the connectivity of any mobile node to the Internet with low latency. In this approach, previous \ac{AP} and new \ac{AP} will coordinate to speed up the re-association process of the mobile device to the new \ac{AP}. Reference~\cite{dimopoulou2005fast} provides a  through review of the existing fast handover strategies for \ac{WLAN} networks.

A common drawback of these approaches is that the prediction of the handover event is only based on the location of the mobile terminal. In a wireless network with directional communications, however, even fixed terminals may need to execute the handover procedure if an obstacle appears on the communication link. When handover becomes frequent, e.g., in \ac{mmWave} networks~\cite{shokri2015mmWavecellular}, soft handover strategies seem to be a better option to support low-latency services. \etal{Semiari}~\cite{Semiari2017Caching} considered a dense network of dual-mode \acp{AP} that integrate both mmWave and microwave frequencies. Each mobile terminal can cache their requested contents by exploiting high capacity of \ac{mmWave} connectivity whenever available, and consequently avoid handover while passing cells with relatively small size. \etal{Xu}~\cite{xu2016distributed} introduced a distributed solution approach that reserves some portions of the capacity of each \ac{AP} to ensure fast handover in \ac{mmWave} networks. In particular, the proposed algorithm maintains two lists for every \ac{AP}: ``active'' and ``potential'' terminals. In the case of handover event for potential terminals, the old and new \ac{AP} coordinate to eliminate the registration latency and support seamless handover.

\subsubsection{Energy-Delay-Spectral Efficiency Trade-offs}
There are applications that impose less stringent latency requirements than URLLC use cases, but require long life-time. Sensors packaged in certain products and sensors implanted in the human body, for example, typically tolerate a couple of hundreds of milliseconds latency,  but require a battery lief-time of years or decades. For such applications, the energy-delay-spectral efficiency trade-off arises as an important issue.
Modern wireless devices are equipped with rate-adaptive capabilities which allow the transmitter to adjust the
transmission rate over time. Associated with a rate, there is a corresponding power expenditure
that is governed by the power-rate function. Specifically,
such a function is a relationship that gives the amount
of transmission power that would be required to transmit at a
certain rate.  For most encoding schemes, the required power is
a convex function of the rate, which implies that transmitting data at a low
rate and over a longer duration has less energy cost compared
to a fast rate transmission. This fact, first observed in \cite{elif}, induces a trade-off between transmission energy and delay.  In particular, in \cite{elif} the problem of rate adaptation subject to deadline constraints on packets was studied, and it was shown that a ``lazy'' packet scheduling scheme that transmits at the lowest possible rate to meet the deadline is energy optimal.  This idea was extended in \cite{zafer} to general quality-of-service (QoS) constraints, where packets have different deadlines, and using techniques from network calculus \cite{cruz}, an energy-optimal rate adaptation policy was devised. Moreover, algorithms for minimum energy transmission over time-varying channels were developed in \cite{zafer2}. A comprehensive overview of energy-delay trade-offs in wireless networks can be found in \cite{berrymodianozafer}.

\subsection{Routing Layer Techniques}\label{sec: routing-intra-net}
To ensure a desired end-to-end latency in a multihop network, the routing policy plays an important role in end-to-end latency. In this subsection, we first focus on the routing algorithms that are proposed for fixed topologies and then discuss those developed for dynamic and ad~hoc networks.

\subsubsection{Routing Algorithms Based on Back-pressure Scheduling}
Throughput optimality is at the heart of most existing routing algorithms. \ac{BP} routing algorithm~\cite{Tassiulas1992} is amongst the most celebrated ones that is provably throughput optimal, but has quadratic end-to-end queuing delay in terms of the number of hops.
When implementing the traditional back-pressure scheduling, each node keeps a separate queue for each flow. Then, in each time slot, each link checks the queues for all the flows passing through it, finds the flow with the maximum differential backlog, and uses this value as the weight for that link. At last, the transmission scheme with the highest weight under the interference constraints will be scheduled.

\etal{Bui}~\cite{Bui2011} modified the \ac{BP} algorithm to make the end-to-end delay grow linearly, instead of quadratically, with the number of hops. The price is a minor throughput loss, which might be of secondary importance for low-latency networks.
Their proposed algorithm uses the concept of shadow queues to allocate service rates to each flow on each link with a substantially lower number of real queues.

Another limitation of the \ac{BP} routing algorithm is that every device in the network should be able to apply the \ac{BP} algorithm. In a real network, however, we may be able to upgrade just a few routers. To address this problem, \etal{Jones}~\cite{Jones2014} considered an overlay architecture in which a subset of the nodes act as overlay nodes and perform the \ac{BP} routing, while the other nodes can perform legacy routing algorithms such as shortest-path~\cite{bellman1958routing,dijkstra1959note}. Simulation results show that with a small portion of overlay nodes, the overlay network can achieve the full throughput, and outperforms the \ac{BP} algorithm in terms of delay.
A decentralized overlay routing algorithm was proposed to satisfy the average end-to-end delay for delay-sensitive traffic~\cite{singh2017optimal}.

In a wireless network, the delay performance of any scheduling policy is largely affected by the mutual interference, caused by the concurrent transmissions.
\etal{Gupta}~\cite{Gupta2009} derived a lower bound for the average delay of a multi-hop wireless network with a fixed route between each source-destination pair. The authors proposed a variant of \ac{BP} to optimize the energy-delay-spectral efficiency trade-offs and to perform close to the delay lower bound.

\subsubsection{Non-back-pressure-based Routing}
The \ac{BP} routing algorithm and its variants utilize the queue length as the metric for the delay~\cite{Tassiulas1992,Bui2011,Jones2014,Gupta2009}. To keep the queues stable and achieve throughput optimality, \ac{BP} algorithms prioritize highly loaded queues. When a flow has light traffic, this queue length metric can lead to a huge delay as a short queue gets a small chance to be served. To address this problem, head-of-line delays can be used to act as the link weights instead of queue lengths~\cite{McKeown1999,Neely2013}. For a multihop network with a contention-based \ac{MAC} protocol,~\cite{DiMarco2015} develops a new metric, called Q-metric, to jointly optimize the routing and \ac{MAC} layer parameters with the objective of optimizing the energy-delay trade-off.

Recently, a new throughput optimal routing and scheduling algorithm, known as \ac{UMW}, was proposed in \cite{sinha17}.  Unlike \ac{BP}, UMW uses source routing, where each packet is routed along a shortest path and the cost of each edge is given by the queue backlog in a corresponding virtual network. Since \ac{UMW} uses source routing, it can choose a ``shortest-path'' for each packet and avoid the looping problem that is inherit to \ac{BP}. Consequently, \ac{UMW} results in a significant delay reduction as compared to \ac{BP} and its variants~\cite{sinha17}.

The surveyed routing algorithms need a fixed and known topology a priori. In an ad~hoc network, however, the topology may be unknown and dynamic, and the cost of topology discovery may not be negligible when we consider low-latency services. In the following, we review some new classes of routing algorithms that suit ad~hoc networks.

\subsubsection{Ad Hoc Network Routing}
The ad-hoc network is representative for scenarios without static structure, like vehicular networks.
The seminal work of~\cite{ElGamal2006} shows how the delay scales with the number of devices in the network.
For static random networks with $n$ nodes,
the optimal throughput per node scales as $T(n) = \Theta(1/\sqrt{n\log n})$, and the delay scales as $D(n) = \Theta(n/\sqrt{\log n})$\footnote{Notation: i) $f(n) = \mathcal{O}(g(n))$ means that there exists a constant $c$ and an integer $N$ such that $f(n) \le c g(n)$ for $n>N$. ii) $f(n)=\Theta(g(n))$ means that $f(n) = \mathcal{O}(g(n))$ and $g(n) = \mathcal{O}(f(n))$.}. The results suggest that there is a trade-off between the throughput and the delay by varying the transmission power, i.e., $T(n) = \Theta(D(n)/n)$. Higher transmission power increases the transmission range and can decrease the delay, due to the possibly reduced number of hops. However, it also increases the interference, leading to a drop in the throughput performance. \etal{Yi}~\cite{yi2003capacity} extended these results to the case of directional antennas and showed that, under ideal beamforming assumptions, the throughput gain scales with $\theta_t^{-1}\theta_r^{-1}$ where $\theta_t$ and $\theta_r$ are the antenna beamwidths of the transmitter and the receiver, receptively. Grossglauser and Tse~\cite{Grossglauser2001} showed that mobility improves transmission range and achievable throughput at the price of higher delay. Specifically, the achievable throughput can be improved to $T(n) = \Theta(1)$, and delay scales as $D(n) = \Theta(n \log n)$. Note that these works do not consider the overhead of channel estimation, nor signaling and computational overheads, so that the actual delay can be larger.

The delay and capacity trade-off for multicast routing with mobility in ad hoc networks is studied in \cite{Wang2011}. The results are based on 2-hop relay algorithms without and with redundancy (redundancy is used to denote transmitting redundant packets through multiple paths with different relay nodes). When a packet is sent to $k$ destinations, the fundamental trade-off ratio is $D(n) / T(n)\ge \mathcal{O}(n \log k)$. The delay of the 2-hop relay algorithm with redundancy is less than that without redundancy, but will cause decrease to the capacity.  More recently, in \cite{talak17} the scaling of capacity and delay for broadcast transmission in a highly mobile cell partitioned network was studied. Using an independent mobility model as in \cite{Neely2005} it was shown that, in a dense wireless network (number of nodes per cell increases with $n$), the broadcast capacity scales as $1/n$, while the delay scales as $\log \log n$.  Surprisingly, it is also shown that both throughput and delay have worse scaling when the network is sparse.

SPEED~\cite{He2002} proposed a routing algorithm to support real-time communications via providing per-hop delay guarantees and bounding the number of hops from source to destination. SPEED is extended in~\cite{Felemban2006} to satisfy multiple QoS levels measured in terms of reliability and latency. To this end, this approach employees a localized geographic packet forwarding augmented with dynamic compensation for local decision inaccuracies as a packet travels towards its destination. Therefore, there is no need for any global information or a centralized entity to design the routing, making this algorithm suitable for large and dynamic wireless sensor networks.

\etal{Zhang}~\cite{zhang2005qos} proposes a variation of the celebrated proactive distance vector routing algorithm~\cite{Perkins99AODV} to guarantee end-to-end latency. In particular, each node establishes and maintains routing tables containing the distance (number of hops) and next hop along the shortest path (measured by the minimal number of hops) to every destination. To support a delay-sensitive service, the transmitter probes the destination along the shortest path to test its suitability. If this path meets the delay constraint, the destination returns an ACK packet to the source, which reserves link-layer resources along the path. Otherwise, the destination initiates a flood search by broadcasting a route-request packet. The flooding is controlled by the delay constraint, namely an intermediate node forwards the route-request packet only if the total delay to the destination is less than the threshold. When a copy of this packet reaches the source with a path that meets the delay constraint, the route discovery process is complete. \cite{Lin2001} proposed a similar idea for mobile networks where the latency requirement will be met by reserving sufficient link-layer resources along the shortest path.

There are several survey papers on the \ac{QoS} routing protocols for general ad~hoc networks~\cite{hanzo2007survey} and for vehicular networks~\cite{paul2012survey,altayeb2013survey} where the topology is dynamic.

\subsection{Transport Layer Techniques}
The transport layer is responsible for the flow and congestion controls, and affects the queueing latency $T_q$. In general, the communication ends may belong to different network operators. However, when they both belong to the same network, we can optimize the transport layer for low-latency services.
\ac{TCP} and \ac{UDP} are the two main protocols in this layer. \ac{TCP} provides reliable end-to-end communications, independent of the underlying physical layer\footnote{Note that packets may traverse several networks and therefore different \ac{PHY} technologies. For example, to watch a YouTube video on our mobile phone, the packets will be transmitted over some fibers (YouTube server to our serving \ac{BS}/\ac{AP}) and also wireless links (\ac{BS}/\ac{AP} to our phone).}, while \ac{UDP} does not have such a reliability guarantee~\cite{Comer2006}. When observing a packet loss at the receiver, \ac{TCP} assumes that the underlying network is congested and intermediate queues are dropping packets. Therefore, it reduces the transmission rate at the transmitter to a small baseline rate to alleviate the congestion in a distributed fashion, of course at the expense of a higher end-to-end delay. This congestion control approach is problematic in the presence of some faulty physical layers (e.g., wireless links) in the end-to-end connection, because by every packet loss, which may happen frequently, \ac{TCP} drops the transmission rate dramatically. Moreover, the measure of \ac{TCP} to detect congestion cannot help early congestion avoidance and leads to the well-known ``full buffer problem''~\cite{floyd1994tcp}. In the last decade, there have been tremendous efforts to address these problems with active queue management, and many of them are surveyed in \cite{Adams2013Active}.

To reduce the end-to-end delay, the existing solutions essentially change the congestion indicators, feedback types, and control functions at every intermediate node. For instance, \ac{CoDel}~\cite{Nichols2012} changes the congestion indicator to a target queue latency experienced by packets in an interval, which is on the order of a worst-case round trip time. \ac{CoDel} starts dropping packets when the expected queue latency exceeds the threshold. The authors showed that \ac{CoDel} absorbs packet bursts in a short term manner, while keeping buffers far from fully occupied in the long term, to guarantee low-latency performance.

To reduce the latency in a mesh network, \etal{Alizadeh}~\cite{Alizadeh2012} propose the \ac{HULL} architecture. The main idea of \ac{HULL} is similar to that of \ac{CoDel}, i.e., keeping the buffers of intermediate nodes largely unoccupied. \ac{HULL} trades bandwidth for buffer space, as low-latency packets require essentially no buffering in the network.
\ac{HULL} uses a counter, called a phantom queue in~\cite{Alizadeh2012}, to simulate queue buildup for a virtual egress link with a slower rate than the actual physical link (e.g., 95\% of the link rate). The counter is put in series with the switch egress port, and is incremented by every new received packet, and decremented according to the virtual drain rate (e.g., 95\% of the link rate). When the counter exceeds the threshold, it will send explicit congestion notifications to adjust the contention window size adaptively. Essentially, \ac{HULL} sends early congestion signals before the saturation of the queues and reserves some portion of the link capacity for latency-sensitive traffics to avoid buffering, and the associated large delays.

\subsection{Transmission Capacity Boosting and Sharing Techniques}\label{sec: capacity-boosting}

In the following, we review some recent techniques that were originally proposed for boosting the transmission capacity, but have the potential to reduce the end-to-end latency.

\subsubsection{MU-MIMO}

\ac{MU-MIMO} and beamforming are essential elements of almost all modern wireless systems including LTE-A and IEEE 802.11ac. Notably, these techniques can help steer the radiated/received energy beams toward the intended locations while minimizing interference, thereby improving the capacity region of the system~\cite{goldsmith2003capacity,spencer2004zero,yoo2006capacity}. We can now serve multiple users in the same time-frequency channel, which can substantially reduce the delay-to-access component of the \ac{MAC} layer latency. However, MU-MIMO
requires the knowledge of \ac{CSI} at the transmitter and at the receiver.

As investigated in~\cite{Marzetta2010,Rusek2013}, increasing the number of antennas simplifies the design of beamforming. In the asymptotic regime where the number of transmit antennas goes to infinity, very simple beamforming schemes, like matched filters~\cite{tse2005fundamentals}, become optimal, as the wireless channels among the transmitter (e.g., \ac{AP} or \ac{BS}) and different devices become quasi-orthogonal~\cite{Rusek2013}.
However, the price of increasing the number of antennas is a higher \ac{CSI} acquisition delay, which may limit the applicability of this technique for low-latency services. This problem is exacerbated in \ac{mmWave} communications, as we discuss in the next subsection.
Beamforming design for \ac{MIMO} networks has a very rich literature with focus on spectral efficiency~\cite{huh2012network, cheung2014spectral, shin2017coordinated}, energy efficiency~\cite{cheung2014spectral, he2015energy, ngo2017total}, and interference cancellation~\cite{zhang2010adaptive, hosseini2014large, huberman2015mimo}, among others. However, surprisingly, designing beamforming for low-latency MIMO networks is a largely open problem.

\subsubsection{Millimeter-wave}
\label{subsubsection: mmWave}
\ac{mmWave} systems operate on a large bandwidth and employ large antenna arrays to support extremely high-data rate services, including the 8 Gbps peak data rate of IEEE 802.11ad and 100~Gbps of IEEE~802.11ay\footnote{Detailed information about this project can be found at \url{http://www.ieee802.org/11/Reports/ng60_update.htm}.} for a single link~\cite{ghasempour2017ieee}. In a multiuser \ac{mmWave} network, the use of large antenna arrays drastically reduces the interference footprint and boosts the throughput, as shown in~\cite{Singh2011Interference,Shokri-Ghadikolaei2015a,Shokri2015Transitional,shokri2016Spectrum,Niu2015Blockage,kinalis2014biased,Boccardi2016Spectrum,petrov2017interference}. As a result, \ac{mmWave} networks experience almost negligible transmission latency, but they may suffer from delay-to-access which includes pilot transmission and beamforming design. \cite{heath2016overview} overviews existing approaches for beamforming design in \ac{mmWave} networks. Most of the existing approaches are based on some iterations (or equivalently a huge number of pilot signals) among transmitters and receivers, which could be very time-consuming given the low transmission rate of the control signals~\cite{Shokri-Ghadikolaei2015}. To reduce the beamforming setup delay,~\cite{Zhang2015} and \cite{zhao2017angle} augmented the beamforming part by a tracking algorithm based on extended Kalman filters, which track the second-order statistics of the channel. However, those approaches need a mobility model. \etal{Olfat}~\cite{Olfat2017Learning} alleviates that assumption by proposing a model-free (data-driven) approach based on reinforcement learning to drastically reduce the number of pilots. Still, the area of low-overhead beamforming design for large antenna arrays is in its infancy and needs further research.

\etal{Ford}~\cite{Ford2016} studied the feasibility of supporting low-latency services in \ac{mmWave} cellular networks, particularly 20 Gbps data rate and 1~ms delay as specified by IMT 2020~\cite{soldani2015horizon}. The authors focused on beamforming aspects, \ac{MAC} layer, congestion control at the transport layer, and core network architecture\footnote{Due to the strong connection to the inter-network latency components, we discuss low-latency core network architectures and edge computing in the next section.}, and proposed a set of solution approaches to meet the delay and throughput requirements. In particular, the authors concluded that digital beamforming using low resolution A/D converters is a good choice for reducing beamforming delay and control channel overhead. Moreover, a flexible transmission time interval exhibits a better latency performance at the \ac{MAC} layer than the conventional fixed transmission time interval.

\subsubsection{Full-duplex}
Full-duplex techniques enable transmission and reception at the same time-frequency resources, leading to a substantial improvement in the transmission capacity and therefore a reduction in the end-to-end latency~\cite{choi2010achieving}. The challenge is the strong self-interference, which can be alleviated by passive suppression (mainly related to antenna design) and active suppression (signal processing and beamforming). The latter increases the processing delay.

The low-latency merits of full duplex are in three domains. First, it decreases the transmission time ($T_t$) by increasing the spectral efficiency. Second, it decreases the \ac{MAC} layer latency, as it reduces the potential contention, especially for the star topology where the central coordinator can send a message to one node while receiving an uplink message from another one at the same time. Last, it can decrease the routing latency ($T_r$). With full duplex, the route selection algorithms may activate adjacent hops simultaneously where an intermediate (i.e., relay) node operates in both downlink and uplink directions, which can substantially reduce the routing delay.

To harvest the gains of full duplex, good cancellation techniques as well as a good \ac{MAC} layer design are necessary and other research problems need to be addressed. The interested readers can refer to \cite{zhang2016full} and the references therein.

\subsubsection{D2D}
\label{subsection:D2D}

\ac{D2D} communication enables direct communication between devices without going through the core of a cellular network~\cite{Li2014a,li2014overview}.
\ac{D2D} communication is a promising solution for the increasing number of connected devices and data rate. It helps to reduce the latency in two perspectives. First, \ac{D2D} enables the devices to communicate with each other in single hop or fewer hops than communication via a \ac{BS}, thus reducing the routing delay $T_r$~\cite{gandotra2016device}. Secondly, the local traffic is separated from the global network (local traffic offloading). Thus, the \ac{D2D} mode reduces medium access delay $T_a$, as the devices in D2D mode retrieving from the local source devices are fewer than those communicating with \ac{BS}, and will access the channel faster.
Non-\ac{D2D} mode devices retrieving contents from the core network also benefits with lower $T_a$, because D2D mode offloads part of the contending devices.

If the \ac{D2D} mode shares the same resources with the core network communication, resource allocation should be used to mitigate interference and guarantee low latency to both \ac{D2D} and non-\ac{D2D} users. The integration of \ac{mmWave} and \ac{D2D} can substantially alleviate the resource allocation problem~\cite{qiao2015enabling}, thanks to the small interference footprint of \ac{mmWave} networks~\cite{Singh2011Interference,Shokri-Ghadikolaei2015a,Shokri2015Transitional,shokri2016Spectrum,Niu2015Blockage,kinalis2014biased,Boccardi2016Spectrum,petrov2017interference}.
\etal{Niu}~\cite{niu2016exploiting} showed that we can add many \ac{D2D} links to a \ac{mmWave} cellular network so as to substantially boost the network capacity. The authors have also developed a simple scheduling scheme to activate concurrent transmissions to support low-latency content downloading. The main challenges are the device discovery and the exchange of control signals, which are usually much harder in \ac{mmWave} networks~\cite{shokri2015mmWavecellular}.

\subsubsection{Cloud RAN and Mobile Fronthaul}
A radio \ac{BS} consists of a \ac{BBU} and a radio frequency unit.
The concept of cloud \ac{RAN} consists in breaking the fixed topology between \acp{BBU} and \acp{RRH}, and to form a virtual \ac{BBU} pool for centralized control and processing~\cite{Chen2011}. Cloud \ac{RAN} supports inter-cell communication and joint processing, which can reduce the routing latency ($T_r$) as well as the processing latency ($T_{\mathrm{pr}}$).
Mobile fronthaul is a novel optical access method that connects a centralized BBU to a number of \acp{RRH} with fiber links in mobile networks~\cite{Chanclou2013}.
A SDN-controlled optical topology-reconfigurable fronthaul architecture is proposed for 5G mobile networks~\cite{Cvijetic2014}. The SDN-controlled fronthaul architecture is responsible for the dynamic configuration of the \acp{BBU} and \acp{RRH} connections to support coordinated multipoint and low-latency inter-cell \ac{D2D} connectivity. The experimental results show that with 10~Gbps peak data rate, sub-millisecond end-to-end \ac{D2D} connectivity is achievable~\cite{Cvijetic2014}.

In \cite{Beyranvand2015}, a combination of fiber and \ac{mmWave} is proposed for efficient fronthauling to lower the cost and support mobility in small cells and moving cells. mmWave is used for transmission between the large number of \acp{RRH} or moving \acp{RRH} and a remote antenna unit, and the fiber optic is used for the connection between the remote antenna unit and BBU pool. To reduce the latency caused by the conversion between optic signal and \ac{mmWave} signal, analog waveform transmission is used for eliminating the digital processing, and a uni-travelling carrier photodiode optical-to-electrical converter is used to provide fast conversion.

\subsection{Cross-layer Techniques}

\subsubsection{Forward Error Correction and Hybrid Automatic Repeat Request Techniques}
\ac{ARQ} is a simple error-control method for data transmission that uses acknowledgements and timeouts to achieve reliable data transmission over an unreliable physical layer~\cite{tanenbaum1996computer}. Acknowledgments are short messages sent by the receiver indicating that it has correctly received a packet, and timeout is a predefined latency allowed to receive an acknowledgment. If the sender does not receive an acknowledgment before the timeout, it usually re-transmits the frame/packet until the sender receives an acknowledgment or exceeds a predefined number of re-transmissions.
Block acknowledgement --initially defined in IEEE 802.11e-- is a simple way to reduce the \ac{MAC} layer latency, especially for high-throughput devices, by sending one acknowledge packet for multiple data packets~\cite{tinnirello2005efficiency}.

\ac{HARQ} combines high-rate forward error-correcting coding and \ac{ARQ} error-control to flexibly perform retransmission of incremental redundancy or a complete new retransmission according to different channel states. If the received packet can not be decoded correctly due to high noise, incremental redundancy can be retransmitted for joint decoding together with the previously received packets. If strong interference is the reason, a new start of transmission is needed.
There is a trade-off between the transmission time $T_t$ and medium access delay $T_a$. With lower coding rate, $T_t$ gets longer, while $T_a$ is shorter as the message can be received successfully with a higher probability and possibly fewer retransmissions.

\subsubsection{MAC-aware Routing Algorithms}

The delay and reliability performance interaction between the \ac{MAC} and routing of the protocol IEEE 802.15.4 is analytically studied in~\cite{DiMarco2015}. Therein, it has been shown that the \ac{MAC} parameters will influence the performance of different routing paths, and, in turn, the traffic distribution determined by the routing will also affects the \ac{MAC} parameters. For a given topology, the \ac{MAC} parameters (affecting $T_a$) can be tuned to satisfy a certain reliability and latency requirement by using the Q-metric proposed in~\cite{DiMarco2015}. The Q-metric measures the contention level without measuring the queues, and adapt the routing patterns (affecting $T_r$). While the back-pressure is proved to be throughput optimal, it is efficient only when the forwarded traffic is high, and it can not capture the contention of low traffic. So Q-metric is more efficient in latency-sensitive wireless sensor networks and other cases where the traffic is low.

A TDMA-based \ac{MAC} for wireless sensor networks that have latency requirements is Delay Guaranteed Routing and \ac{MAC}~\cite{Shanti2010}. However, unlike traditional TDMA \acp{MAC} that require a separate routing mechanism, the routes of Delay Guaranteed Routing and \ac{MAC} as well as the medium access slot schedule are determined and fixed according to the position of each node.
The algorithm is TDMA-based and no retransmissions are permitted. As long as the transmission interval of two successive packets of each node is larger than the TDMA superframe duration, a deterministic upper delay bound and minimal packet loss can be guaranteed.

\subsection{Network Function Virtualization and Software-defined Networking }
\label{subsubsection:NFV-SDN}
\ac{NFV} and \ac{SDN} cannot reduce the latency by themselves alone, but they are the premise of some algorithms that help achieve low latency, so we also study them together as an enabling technique.

\ac{NFV} can virtualize the network node functions into building blocks that may connect, or chain together, to create communication services. With the necessary hardware support, the system can change flexibly by employing different \ac{NFV} blocks. For instance, different asynchronous waveforms can be generated by combining different filters and modulation schemes~\cite{Han2016}, which affects the transmission time ($T_t$).

Though \ac{NFV} can act alone, it is generally working together with \ac{SDN}. \ac{SDN} decouples the control plane and data plane, promoting centralized network control and the ability to program the network~\cite{Kreutz2015}. SDN can be used to implement and manage the NFV infrastructure by combining and tuning parameters of multiple \ac{NFV} blocks, such as to enable flexible configuration of the virtual network slices for different latency priorities~\cite{Ford2016}. By tuning the parameters of different \ac{NFV} blocks, which further virtualize different techniques and affects different latency components, \ac{SDN} and \ac{NFV} may have an ample effect on various latency components.
On the other hand, \ac{SDN} will introduce overheads from the control plane and flow setup latency as it is flow based. Software-based \ac{NFV} also tends to increase the processing latency ($T_{\mathrm{pr}}$) compared to the pure hardware operation. The effect of these on the delay performance should be carefully controlled and minimized.

The ETSI \ac{NFV} architecture~\cite{nfv2network} acts as a reference standard, whose performance and security have been well studied~\cite{virtualisation2014nfv,briscoe2014network}.
\cite{siddiqui2016hierarchical} proposes a \ac{5G} architecture for low-latency and secure applications based on ETSI NFV architecture. To achieve low latency with shared resources, the architecture employs the scalability enabled by \ac{SDN} and \ac{NFV} to perform on-demand caching and switching, which can guarantee low medium access latency ($T_a$), routing latency ($T_r$), and queuing latency ($T_q$) by flexibly allocating different resources based on different traffic load and requirements including latency. Moreover, it also proposes to use a smart network interface card with NFV acceleration capability to mitigate the processing latency ($T_{\mathrm{pr}}$) caused by NFV.
A backbone network that provides high-performance connectivity in Japan is equipped with the capabilities of \ac{NFV} and \ac{SDN}~\cite{kurimoto2017sinet5}.
The \ac{NFV} orchestrator creates virtual network appliances, such as virtual routers, virtual firewalls, and virtual load balancers.
\ac{SDN} enables users to establish connections with specified bandwidth and demand, which flexibly scale up or down the virtual network appliances. In this way, \ac{SDN} affects all the latency components whose parameters has been dynamically tuned, such as transmission time ($T_t$), medium access latency ($T_a$) and routing latency ($T_r$).
The latency is reduced by 20\% compared to the previous version network, and the auto-healing time of a virtual network functions also drops to 30 s from 6 min from the previous version.

\section{Inter-network Techniques and Technologies}\label{sec: Inter-network}
Many modern end-to-end services require packets to traverse multiple networks, usually handled by different operators. Unfortunately, in most practical scenarios, the exact latencies of the intermediate networks are not known, and the service provider may have access only to the delay probability distributions. As a result, controlling the latency in a multi-domain scenario is significantly more complicated than managing latency in single domain scenarios.

In this section, we address techniques that can be used to control inter-network (also called multi-domain) latency. We first review some multi-domain routing algorithms that guarantee end-to-end latency, though they may not meet the tight requirements of low-latency services. We then highlight the importance of content placement and edge caching to facilitate multi-domain routing and substantially reduce latency.

\subsection{Inter-domain Routing}\label{sec: routing-inter-net}
In Section~\ref{sec: routing-intra-net}, we have discussed various routing techniques that can reduce the latency within a domain. \ac{BGP} is a celebrated inter-domain routing protocol used throughout the Internet to exchange routing information among different domains~\cite{rekhter2005border}. In \ac{BGP}, the edge routers frequently send path vector messages to advertise the reachability of networks. Each edge router that receives a path vector update message should verify the advertised path according to the policy of its domain. If it complies with its policy, the router modifies its routing table, adds itself to the path vector message, and sends the new message to the neighbor domain. As a result, edge routers of every domain maintain only one path per destination.

Although \ac{BGP} has evolved for many years, its current implementation is based on the number of hops. Reference~\cite{arins2014latency} proposes a method to measure the latency, share it with the edge routers of the neighboring domains, and modify the \ac{BGP} routing decisions throughout the entire network. \cite{egilmez2012distributed} and \cite{lin2015west} generalize this idea to incorporate a logical \ac{SDN} in a multi-domain network. \cite{egilmez2012distributed} formulated a constrained shortest path problem to minimize a convex combination of packet loss and jitter subject to an end-to-end latency constraint. \cite{paolucci2016interoperable} demonstrated a multi-domain \ac{SDN} orchestrator to select the shortest routes based on end-to-end latency, where the delay statistics are captured by the proposed segment routing monitoring system.

Reference \cite{King2004Routing} extends \ac{BGP} to allow for path-diversity in the sense that multiple paths are advertised by any edge router, multiple QoS metrics including latency are propagated in the \ac{BGP} update message, and multiple routes for any source-destination pair are selected. The authors showed that path-diversity substantially improves load balancing throughout the multi-domain network and can reduce the end-to-end latency. \cite{2017arXiv170307419S} developed a dynamic routing policy for delay-sensitive traffics in an overlay network. In particular, the authors replaced the average end-to-end latency requirement by an upper bound on the average queue length of every flow on every link. Then, considering an underlying legacy network whose latency characteristics are unknown to the overlay network, they formulated a constrained Markov decision process that keeps the average queue lengths bounded, and proposed a distributed algorithm to solve that problem. Although this paper targets a single domain, the main idea can be extended to the case of multi-domain networks. Classical stochastic shortest path \cite{ji2005models} and its online variation \cite{talebi2017stochastic} are highly relevant to the problem of routing design in multi-domain networks when only imperfect knowledge of latency within each domain is available. Multi-domain routing design with limited domain-specific information is a very interesting and wide open area for future research.

\subsection{Edge Caching and Content Placement}
\label{subsection:edge-caching}

Define the feasible latency region as the set of all possible latency values that can be achieved by some routing policy. In the previous subsection, we observed that optimizing inter-domain routing reduces the end-to-end latency. However, when the content is located far from the terminal, the feasible latency region may not include the desired value of the low-latency service; namely, there is no routing algorithm that can lead to the target latency value. As a simple example, consider a line network of 10 domains (including source and destination ones), and assume that each adds at least 10 ms latency. Therefore, the round-trip latency is lower-bounded by 200~ms, which may be way beyond the tolerable latency. Caching popular contents at the edge routers of local domains is a promising approach to bring the contents closer to the devices so as to improve the feasible latency region.

Edge caching brings the content closer to the end terminals, which substantially reduces the inter-domain routing delay. From the local server perspective, fewer terminals (only local ones) contend to access that content, which improves the service rate to those terminals. Moreover, this technique offloads parts of the traffics of the main server, freeing the capacity for other terminals and further reducing the latency. Altogether, edge caching is especially beneficial for services with stringent latency requirements.

The design of efficient caching strategies involves a broad range of problems, such as accurate prediction of demands, intelligent content placement, and optimal dimensioning of caches. Moreover, as the caches are physically scattered across the network and the user requests are generated almost everywhere in the network, caching policy favors low-complexity distributed algorithms.
\etal{Borst}\cite{borst2010distributed} propose a light-weight cooperative content placement algorithm that maximizes the traffic volume served from caches and thus minimizes the bandwidth cost. Maddah-Ali and Niesen~\cite{Maddah2014Fundamentals,maddah2015decentralized} focus on the problem of caching within a network where popular content are pre-fetched into the end-user memories to bypass a shared link to the server and showed that there is a trade-off among the local cache size (i.e., the memory available at each individual user), aggregated global cache size (i.e., the cumulative memory available at all users), and the achievable rate (and latency) of individual terminals. The authors develop a simple coded caching scheme in~\cite{Maddah2014Fundamentals} and its decentralized variant in~\cite{maddah2015decentralized} that substantially improves the memory-rate trade-off. The analysis and proposed algorithms, however, are limited to the single-domain single-server case.
\etal{Doan} propose a novel popularity predicting–caching procedure for backhaul offloading in cellular network, and an optimal cache placement policy to minimize the backhaul load is computed by taking both published and unpublished videos as input~\cite{doan2018content}.

Reference~\cite{tan2013optimal} focuses on a video-on-demand application over a network with two modes of operations: peer-to-peer and data center. In particular, video requests are first submitted to the peer-to-peer system; if they are accepted, uplink bandwidth is used to serve them at the video streaming rate (potentially via parallel substreams from different peers). They are rejected if their acceptance would require the disruption of an on-going request service. Rejected requests are then handled by the data center. The authors developed a probabilistic content caching strategy that enables downloaders to maximally use the peers’ uplink bandwidth, and hence maximally offload the servers in the data centers. \etal{Golrezaei}~\cite{Golrezaei2012} extended the model of~\cite{tan2013optimal} to the scenario of video-on-demand streaming to mobile terminals from Internet-based servers and proposed a distributed caching network to reduce the download latency. In this setting, local caches with low-rate backhaul but high storage capacity store popular video files. If the file is not available in the local cache, it will be transmitted by the cellular network. The authors analyzed the optimal assignment of files to the caches in order to minimize the expected downloading time for files. They showed that caching the coded data can substantially improve the performance in terms of both computational complexity and aggregated storage requirement.
\etal{Bastug}~\cite{Bastug2014} proposed a caching strategy that predicts the demand pattern by the users and caches them, in a proactive manner, in local \ac{BS} during the off-peak hours. When the users actually make the demands, the contents can be retrieved with high probability directly from the cache instead of waiting for the backhaul network and inter-domain routing latency.

\etal{Bhattacharjee}\cite{bhattacharjee1998self} proposed a self-organizing cache  scheme in which every router of the network maintains a small cache and applies an active caching  management strategy to organize the cache contents. \cite{li1999optimal} focused on the optimal placement of $M$ web proxies in $N$ potential sites with the objective of minimizing the overall latency of searching a target web server for a given network topology. The authors formulated this problem as a dynamic program, and obtained the optimal solution in polynomial time.

\subsection{Fog Computing}
\label{subsection:fog-computing}
With \ac{IoT}, billions of previously unconnected devices are generating more than two exabytes of data each day, and 50 billion devices are estimated to connected to the Internet~\cite{fogreview}.
Such large amount of data cannot be processed fast enough by the cloud.
Fog computing is proposed to extend cloud RAN further to the edge, such that any device with computing, storage, and network connectivity can be a fog node~\cite{bonomi2012fog,bonomi2014fog}.
The cloud and fog nodes merge into a new entity, referred to as cloud+fog~\cite{Li2014}.
In a cloud+fog architecture, critical \ac{IoT} data with stringent latency requirement can be processed at the closest fog node to minimize latency, while less delay-sensitive data can be passed to the aggregation node or the cloud.
Fog computing also offloads gigabytes of network traffic from the core network.
Similar to edge caching, fog computing reduces the routing latency and channel establishment latency to both data processed by the fog nodes and by the core network.

Two main challenges in realizing fog computing's full potential are to balance load distribution between fog and cloud, and to integrate heterogeneous devices into a common computing platform~\cite{bessis2014big}.
To evaluate resource-management and scheduling policies across fog and cloud resources, an open source simulator called iFogSim is developed, which can model and simulate the performance in terms of latency, energy consumption, network congestion and operational costs~\cite{gupta2017ifogsim}.
An architecture of Smart Gateway with Fog Computing is presented in~\cite{aazam2014fog}, where the Smart Gateway can collect, preprocess, filter, reconstruct, and only upload necessary data to the cloud. To handle heterogeneous data collected from heterogeneous devices, transcoding and interoperability are either achieved by equipping the Smart Gateway with more intelligence or through the fog computing resources.

\section{Standards for Low-latency and Ultra-reliable Communications}
\label{Sec:Standards}

In this section, we analyze how intra networks and inter network can be supported by current and emerging communication standards. We will review the low-latency characteristics of standards in cellular networks, industrial communication, and WLAN group. One or a combination of these standards are used for the uses cases discussed in Section~\ref{Sec:Requirements}.

\subsection{The 3GPP New Radio and 5G Initiative}
\ac{5G} of mobile communication aims to support 1 ms latency, 10 Gbps peak speed, and, at a global level, 100 billion connections.
\ac{5G} has three main classes of use cases: \ac{URLLC}, enhanced mobile broadband, and massive machine type communication. \ac{URLLC} has the most stringent requirement for very low latency and high reliability, which suits applications such as factory automation, smart grid, and intelligent transportation.
With a vision to provide a unified infrastructure for different use cases, 5G will include both the evolution of 4G and a new radio access technology.

\ac{5G} is designed to operate in a wide range of spectrum including frequency bands below 1~GHz up to 100~GHz. The wide bandwidth at higher frequencies including \ac{mmWave} band can effectively boost the transmission rate and reduce the transmission time ($T_t$). Moreover, flexible deployment of more micro/pico sites at traffic dense spots eases the medium access pressure, and reduces medium access time ($T_a$).
At the physical layer, 5G supports various modulations from QPSK, 16 QAM to 1024 QAM to support different transmission rates and transmission time ($T_t$) for different use cases.
The current candidate waveform is OFDM with scalable numerology and adjustable sub-carrier spacing, CP duration and OFDM symbol duration, which supports adjustable transmission time ($T_t$). A detailed assessment of OFDM in \ac{5G} can be found in~\cite{zaidi2016waveform}.
MAC scheduling in \ac{5G} follows the time-slotted framework of 4G, and the transmission can only start at the beginning of the scheduled slot. To improve the access efficiency of short packets for \ac{URLLC}, the time slot in 4G can be divided into multiple mini-slots (the length can be as short as one OFDM symbol) to enable lower delay to access, which can greatly reduce the medium access time ($T_a$).
Multiple antennas will also be supported in \ac{5G} for MU-MIMO, which also helps reduce the medium access time ($T_a$). With the increase of the carrier frequency, the number of antennas and the multiplexing order will increase. Meanwhile, the complexity to obtain CSI for beamforming also increases. A highly flexible but unified CSI framework is supported by \ac{5G}, which enables different antenna deployments corresponding to different CSI settings.

\subsection{Industrial Communication Networks}
Industry 4.0 is currently seen as the most advanced industrial automation trend, where one of the important aspects is real-time communication between industrial modules. Wireless networks offer simple deployment, mobility and low cost, and are gaining popularity in the industrial sites~\cite{mumtaz2017massive}. The requirements in industrial applications to support high reliability, high data rates and  low latency pose difficulties to wireless networks deployment, where the bottleneck mostly lies in the latency.
To address these challenges, there have been some proposals, such as WirelessHP and IEEE 802.15.4e.

The recently proposed Wireless High Performance (Wireless HP) aims to provide a physical layer solution to support multi-Gbps aggregate data rate, very high reliability level ranging from $10^{-6}$ to $10^{-9}$, and packet transmission time ($T_t$) lower than 1 $\mu$s~\cite{luvisotto2017physical}.
Taking the advantage of deterministic and periodic traffic in the latency-sensitive industrial applications, WirelessHP reduces the PHY layer preamble length (part of $T_h$) (while still ensuring reliable packet decoding) and optimizes OFDM parameters to reduce the inefficiencies that affect short packet transmission.

IEEE 802.15.4 is a successful protocol that also forms the basis for the first open standard WirelessHart for process automation applications in the industrial field. However, the drawbacks of low reliability and unbounded packet latency limit its deployment in the industrial applications that have stringent requirements for latency and reliability~\cite{de2016ieee}. To overcome these limitations, the recently released IEEE 802.15.4e amendment introduces MAC layer enhancements in three different MAC modes. Despite the individual features among the different modes, here we focus on the modifications in terms of latency performance improvement. To guarantee bounded medium access time ($T_a$), channel access is time slotted and included both contention-free and contention-based modes for periodic and aperiodic traffic respectively. The number of time slots can be flexibly tuned according to the traffic load, which avoids the inefficiency of idle slots incurred in fixed framing structures. Channel hopping is applied to combat fading and improve reliability, which in turn will lower the retransmission latency, thus reducing medium access time ($T_a$).

\subsection{IEEE WLAN Group Standardization}
IEEE 802.11 is a set of MAC and PHY specifications to implement \ac{WLAN} in the 0.9, 2.4, 3.6, 5, and 60 GHz frequency bands. The key parameter index of the IEEE 802.11 family focus on data rate, coverage range, connectivity. Though latency performance is not specified in the protocols, the protocol IEEE 802.11ak is designed to support industrial control equipment, and latency should be bounded in this scenario. Moreover, the ability and infrastructure of IEEE 802.11 to support high data rate make it an indispensable part in the ecosystem to achieve  low latency. IEEE 802.11ax, which is due to be publicly released in 2019, is designed to improve spectral efficiency at 2.4 GHz and 5 GHz. The technical highlights are the modulation support of up to 1024 QAM and multiuser support in both frequency and spatial domains by the combination of OFDMA and MU-MIMO, which are effective to decrease the transmission time $T_t$ and medium access latency $T_a$ respectively.
IEEE 802.11ay is the follow-up of 802.11ad working at 60 GHz. Compared to 802.11ad with 2.16 GHz bandwidth, 802.11ay has four times the bandwidth by channel bonding. Moreover, MIMO is added with a maximum of 4 streams with a per-stream link rate of 44 Gbps, which can substantially decrease the transmission time $T_t$ for heavy traffic by using the large bandwidth at higher frequency band.

\subsection{Other Standardization Activities}

Among the whole wide spectrum, only a small portion is regulated as licensed spectrum. While the licensed spectrum has better performance due to less interference, the increasing number of connected devices drives the necessity to use unlicensed spectrum. \ac{LAA}, \ac{LTE-U} and Multefire~\cite{multefire2015lte} are three representatives to explore LTE services in the unlicensed 5 GHz band~\cite{labib2017extending}.
\ac{LAA} and \ac{LTE-U} use the unlicensed band by offloading traffic to boost data rate and reduce transmission time ($T_t$), while the control signals stay in the licensed band.
Multefire takes a step even further, with no anchor at the licensed band, and both the control signal and the data traffic are transmitted in the unlicensed band. Thus Multefire not only helps to reduce the transmission time ($T_t$) thanks to the boosted data rate, but it also reduces the access latency ($T_a$) with more access resources.

IEEE 802 \ac{TSN} aims to deliver deterministic latency over Ethernet networks~\cite{TTTech2015IEEE}. Possible applications include converged networks with real-time Audio/Video Streaming and real-time control streams which are used in automotive or industrial control facilities.
The objective latency for short messages per hop is set to be 4 $\mu$s or less with 1 Gbps transmission rate.
To guarantee deterministic delay for the time-sensitive data, the switches are used to schedule the data transmission.
The switches should be aware of the cycle time of these latency-sensitive data, and during the window expected for the arrival of time-sensitive data, the switch will block non-time-sensitive interfering traffic to eliminate queueing latency $T_q$. To enhance the reliability, more than one path is used simultaneously.

\section{Further Discussions}
\label{Sec:Discussion}

\subsection{Short Packets}
In Section~\ref{Subsubsection:short-packets}, we described research topic about sending short packets when the sizes of payload and header are comparable, how to jointly code the payload and meta data in an efficient way is of great importance to improve the spectral efficiency and to reduce the transmission time $T_t$.
Moreover, for short packets, channel establishment delay $T_a$ can be much longer than the transmission time, which is very inefficient, so the MAC control overhead should be modified.
Another open issue is to get the optimal value of the maximum coding rate for finite packet length and finite packet error probability,
which is an NP-hard problem with exponential complexity~\cite{Durisi2015}.

\subsection{The Trade-off for Low latency}

In Subsection~\ref{Subsection:MAC} and \ref{sec: routing-intra-net}, we described  the research activities around the combinations of different techniques and different parameter settings have different latency performance and other performance indicators. Expectedly, there are trade-offs between the latency performance and other performance metric such as throughput, reliability and energy consumption. There is a trade-off between latency and throughput for different MAC scheduling schemes~\cite{Modiano2009}, the trade-off between latency and reliability of slotted ALOHA and CSMA is shown in~\cite{Yang2003}, and the trade-off between delay and energy consumption in~\cite{elif}. The trade-off between latency and throughput in ad hoc network routing is given in~\cite{ElGamal2006,Grossglauser2001}. The trade-offs may be different for different regions (i.e., different throughput or reliability). When the performance other than latency does not exceed a boundary, the latency grows mildly, while the latency increases sharply when exceeding that boundary.
For different techniques,  such trade-offs are expected to exist, and the related boundaries should be well determined.

\subsection{mmWave}

In Section \ref{subsubsection: mmWave}, we discussed about
mmWave as an enabling technology for low-latency communication, and it is also a research direction that attracts much attention with many open problems to be investigated.

Firstly, beamforming is the premise to support the directional transmission and the very high data rate to help reduce latency, however, the delay caused by beamforming may also prohibit mmWave techniques to work in low-latency communications. Moreover, as the coherence time in mmWave band is shorter than that in microwave band, efficient beamforming algorithms in mmWave prove to be very prominent research direction in mmWave and are essential for mmWave techniques to be used in low-latency scenarios. \etal{Giordani} surveyed some existing beamforming and beam-tracking techniques for 3GPP New Radio at the \ac{mmWave} frequencies~\cite{giordani2018tutorial}.

Secondly, although mmWave communication is mostly noise limited, the SINR at the receiver side might probably degrade  greatly when the beam at the receiver side is also aligned to undesired transmitting beams besides the desired one. Thus, multilevel HARQ is needed to indicate retransmission for certain frames corrupted by random strong noise, or a new transmission when experiencing high interference~\cite{Levanen2014}. When the high interference lasts for a long time, retransmission may not be useful, so effective and efficient solutions are needed.

Another problem occurs in the case of blockage in the main beam aligned link. As the PHY layer frame duration is much smaller compared to the time for the vehicle or people to move away, under this case the device should fall back to search other mmWave BSs in the local area instead of waiting. And if no suitable mmWave BSs are available, UE may also fall back to use microwave. Then whether to reassociate in the mmWave band or to change to microwave band during blockage is an open problem.
An efficient way to determine the reason for performance degradation is of practical importance to reduce the delay in mmWave.

\subsection{Combination with PHY Parameters}

In Subsection~\ref{Subsection:MAC} and \ref{sec: routing-intra-net}, we discussed the effect of PHY parameters on the corresponding latency components.
For contention-based MACs, the probability of a successful transmission is often calculated by the probability when there are no more than one node transmitting at the same time. This practice holds for the omnidirectional transmission. When we use directional transmission with a large number of antenna elements, e.g., in \ac{mmWave} systems, some simultaneous transmissions may not cause interference to each other. Due to this reason, the contention-based MACs need to be revised for certain scenarios.
Moreover, given a reliability requirement, retransmissions due to poor SINR may be needed besides collision during the access phase, so the PHY parameters should be combined to determine the MAC parameters, and more effective and efficient cross layer design to achieve low-latency need to be investigated.

At the routing layer, as the topologies are often unknown and may change in time, it is difficult to determine the efficient interference model. However, in directional communications, multiple links can be activated without causing  interference, and this suggests new research topics at the routing level. The trade-off between the time to establish  directional communication and the improvement of the routing delay should be investigated.

\subsection{Unified Communication Network for Low-Latency Applications}

In Sections \ref{Sec:Standards} and \ref{subsubsection:NFV-SDN}, we described the standardization activities, NFV and SDN separately, here we will discuss the open research that arise under the vision of 5G.
5G aims at providing a unified infrastructure for a wide variety of use cases from media sharing (requiring high transmission date), massive machine type communication (requiring access  availability), to applications that requires low latency. Moreover, for the latency-sensitive applications, different requirements in terms of latency, reliability, transmission rate and energy consumption are all different. When considering different applications as verticals, they operate on top of horizontal communication infrastructures, resource, and techniques~\cite{wollschlaeger2017future}. A unified communication ecosystem should figure out how to share horizontal infrastructures and resource tailored to different verticals. NFV and SDN are necessary to enable flexible configuration of the communication services. When designing protocols, interfaces between different communities should also be taken into consideration to ensure easy and efficient combination of the horizontal techniques.

Traffic offloading is another key technique to promote a unified communication network.
By 2020, 5G is expected to increase the area capacity 1000-fold, and to be able to connect 100 billion devices. Despite the remarkable growth in capacity, it is still difficult to support such a large number of connections. Even with different priorities, latency-sensitive transmission may not be guaranteed.
Traffic offloading may greatly relieve the burden of the mobile network by using other communication forms to share the traffic. For instance, local traffic can be directly exchanged by D2D communication mode without going through the infrastructure.
An architecture unifying the coverage-centric 4G mobile networks and data-centric fiber-wireless net broadband networks was proposed in~\cite{Beyranvand2015}. This architecture uses fiber as backhaul sharing and WiFi to offload the mobile user traffic from the 4G mobile networks,
and as a result the end-to-end delay decreases dramatically on the order of 1~ms.

\subsection{Age-of-Information}
\ac{AoI} is a new performance metric that measures the amount of time that elapsed since the most recently received packet was generated at its source. As such, \ac{AoI} measures the ''freshness of information" from the perspective of the destination. \ac{AoI} has been receiving increasing attention in the literature, particularly for applications that generate time-sensitive data such as position, command and control, or sensor data. \ac{AoI} is a function of packet delay and packet inter-delivery time. Thus, low delay alone many not be sufficient to achieve good \ac{AoI} performance. For example, an M/M/1 queue with a low arrival rate and a high service rate may have low queueing delay but high \ac{AoI} because the packet inter-arrival times are large. Thus, achieving good \ac{AoI} performance involves a balance between maintaining low delays and small packet inter-arrival times.  To better understand
these phenomena, reference \cite{yates} modeled the network between the source
and destination as a single first-in-first-out (FIFO) queue, and proved that there is indeed an optimal rate at which \ac{AoI} is
minimized.
Since then, most of the work on \ac{AoI} has focused on
single queue models. Age for FIFO M/M/1, M/D/1, and
D/M/1 queues was analyzed in \cite{yates}, multiclass FIFO M/G/1
and G/G/1 queues were studied in \cite{longbo}.  However, the problem of minimizing \ac{AoI} in a wireless network with interference constraints has received limited attention.  In \cite{igor16}, the authors develop scheduling algorithms for minimizing \ac{AoI} over a wireless base-station, where only one node can transmit at a time, and in \cite{rajat18} the authors study the \ac{AoI} minimization problem in a wireless network subject to general interference constraints. The approach is generalized to multi-hop wireless networks in \cite{rajat17b}.

\subsection{Hardware and Smart Devices}

Due to the limitation of our expertise, we mainly discuss from technique respective, there are also many open research problems about hardware and smart devices which enable the functioning of different techniques.
When we seek to exploit the wide bandwidth at higher frequencies such as mmWave, the according hardware that operates in the high bands are needed, and technologies in the higher frequencies can be widely used to achieve good performance when the price of the hardware becomes lower.

Nowadays the devices are becoming more and more smart in terms of better processing capability and larger caching size. This smartness not only enables installing and running applications smoothly, but also help to achieve low-latency.
With larger caching capability, the devices can proactively cache popular contents by an analysis of the user preference from the previous contents the user has browsed \cite{Bastug2014}. Then when the user actually makes the request to these cached contents, the device can display them with local cache instead of accessing through the network.

Another direction is to enable the devices to support D2D communication from both the hardware and protocol perspective \cite{Boccardi2014}. Then the devices can communicate with others in proximity using lower power and probably in a single hop, instead of communicating through the BS going through multiple hops or using strong transmitting power. However, the trade-off between the extra overhead for control and channel estimation and the delay saved by communicating in the D2D mode still needs further research.

\subsection{Security}
Currently, the packet overheads due to security or privacy methods substantially contribute to the delay both for the communication of information itself and for the decoding/processing. The classic approach to privacy and security is cryptography~\cite{goldreich2009foundations}. However, cryptography introduces formidable overheads and
heavy coordination over the transmitters/receivers that are involved, which substantially makes it impossible to achieve very low latencies.
Existing alternative privacy methods for low latency networking demand substantial investigation. Alternative
methods to cryptography – namely, information-theoretic secrecy~\cite{harrison2013coding}, differential privacy~\cite{dwork2014algorithmic}, k-anonymity~\cite{sweeney2002k}, and signal processing security methods~\cite{sankar2013role} – present shortcomings when it comes to their use for low latency networking. Like cryptography,
information-theoretic secrecy prevents an eavesdropper acquiring information from two communicating
nodes. However, the channel acquisition phase that is needed for these methods includes substantial delays that are in contrast with low latencies. Differential privacy, k-anonymity and signal processing methods
perturb the original data to make data analysis, and as such it is not clear how they can be used to ensure private or secure low latency communications.

\section{Conclusion}
\label{Sec:Conc}

Low-latency communications are arguably the most important  direction in the next generation of communication networks. In this paper, we showed that the most  prominent use cases demanding low latency are supported by complex network interactions, including inter-network and intra-network interactions. To realize low-latency networks, it is important to determine where and how latency occurs and what methods can help to reduce it. We investigated how the delay accumulates from physical layer to transport layer, and we showed how to characterize the end-to-end delay into several components. Then we discussed how different techniques may influence one or multiple delay components. We argued that these techniques should be optimized together to reduce the delay while satisfying other requirements such as reliability and throughput. \ac{MU-MIMO}, mmWave, and full-duplex, which can greatly improve the data rate, were considered as three enabling technologies to support low-latency communication, but each of the three also poses challenges, e.g., \ac{MU-MIMO} and mmWave will introduce beamforming delay before the start of the transmission. Finally, sending short packets, the trade-off between latency and other network performance indexes, using mmWave bands, design with physical parameters, hardware and smart devices design, and traffic offloading are some of the most promising research areas that will need substantial future research developments.

\bibliographystyle{IEEEtran}

\bibliography{references}

\end{document}